\documentclass{styl/aa}

\usepackage{xcolor}                 
\usepackage{CJKutf8}                
\usepackage{enumitem}               
\usepackage[
    colorlinks=true, 
    citecolor=blue, 
    linkcolor=blue, 
    urlcolor=blue
]{hyperref}                         

\defcitealias{eriksson.LEJ.2026.07}{E26}

\begin{document}

\title{GameDev: GPU-accelerated model for dust evolution}
\titlerunning{the GameDev code}

\author{Jiaqing Bi \begin{CJK*}{UTF8}{gbsn}(毕嘉擎)\end{CJK*}}
\authorrunning{Jiaqing Bi}

\institute{Max-Planck-Institut für Astronomie, 69117 Heidelberg, Germany; \texttt{bi@mpia.de}}

\date{Manuscript compiled on \today}

\abstract{
    Grain growth and fragmentation through dust collisions in protoplanetary disks are strongly coupled with dust dynamics. This is because the spatial and velocity distributions of dust set the collision rates and outcomes, while dust grain sizes set the aerodynamic coupling with gas. However, existing dust evolution codes often rely on prescribed dust velocities and are mostly restricted to one or two spatial dimensions. In this work, we present \texttt{GameDev}, a dust evolution model accelerated by graphics processing units (GPUs) that couples independently evolved three-dimensional dust dynamics with Monte Carlo collisional evolution of dust. In \texttt{GameDev}, we model dust as Lagrangian representative particles in a prescribed gas disk. We integrate particle trajectories with the staggered semi-analytic method, which remains accurate for both tightly and weakly coupled dust grains. We estimate collision rates from each particle's nearest neighbors rather than on a grid, preventing physically close particles from being excluded from collision sampling because they lie across a grid boundary. Most importantly, we sample collision events against neighborhoods that remain frozen over adaptive intervals, enabling massive parallelization across particles. These implementations make \texttt{GameDev} an accurate and efficient tool for studying dust evolution in protoplanetary disks. \texttt{GameDev} supports both NVIDIA CUDA and AMD ROCm platforms, provides a smaller-scale standalone Eulerian dust fluid model, and is publicly available on GitHub.
}

\maketitle
\nolinenumbers


\section{Introduction}

Dust in protoplanetary disks plays a central role in forming planets. In particular, grain growth from the micron sizes preserved in the interstellar medium to the millimeter sizes observed in protoplanetary disks represents a critical first step (\citealt{testi.L.2014.01,drazkowska.J.2023.07,birnstiel.T.2024.09}, and the references therein). Numerous laboratory experiments and numerical models suggest that this initial grain growth is mainly facilitated by coagulation through collisions (\citealt{dominik.C.1997.05,blum.J.2000.01,dullemond.CP.2005.05,blum.J.2008.09}). Meanwhile, dust collisions not only result in sticking that drives coagulation, but can also lead to other outcomes such as bouncing, partial mass transfer, erosion, and fragmentation (\citealt{guttler.C.2010.04,zsom.A.2010.04,schrapler.R.2011.06,krijt.S.2015.02,blum.J.2018.03}).

Besides the direct control of the grain-size distribution, collisions also impact dust dynamics through the grain-size-dependent aerodynamic coupling with gas (\citealt{weidenschilling.SJ.1977.07}). Consequently, the radial drift, vertical settling, turbulent mixing, and clumping are all closely associated with the local collisional activity (\citealt{adachi.I.1976.12,nakagawa.Y.1986.09,cuzzi.JN.1993.11,dubrulle.B.1995.04,takeuchi.T.2002.12,dullemond.CP.2004.07,paardekooper.SJ.2004.10,youdin.AN.2005.02,birnstiel.T.2009.08,squire.J.2018.07}). Meanwhile, the spatial distribution and velocity of dust, primarily controlled by its dynamics, can in turn impact collision rates and outcomes as well (\citealt{volk.HJ.1980.05,markiewicz.WJ.1991.02,dullemond.CP.2005.05,ormel.CW.2007.05,brauer.F.2008.03,windmark.F.2012.08,booth.RA.2016.05,tominaga.RT.2025.04}). Therefore, a closed feedback loop is formed between dust collisions and dynamics, requiring both processes to be resolved self-consistently and concurrently in theoretical and numerical investigations.

Combining dust collisions and dynamics is important not only for theoretical understanding but also for observations. This is because protoplanetary-disk observations depend both on the spatial distribution of dust shaped by its dynamics and on dust opacity, which collisions can substantially alter through changes in grain size, grain shape, porosity, and the chemical composition of coating materials (\citealt{semenov.D.2003.11,draine.BT.2006.01,weidling.R.2009.05,paszun.D.2009.11,ormel.CW.2011.08,kataoka.A.2014.08,tazaki.R.2018.06,krijt.S.2024.11}).

Capturing the interplay between dust collisions and dynamics requires, at a minimum, the evolution of the position, velocity, and polydisperse grain-size distribution\footnote{A polydisperse distribution means that the local dust population is divided into multiple groups and represented by more than one characteristic grain size.} of dust. There are three typical methods to account for these coupled processes in planet formation studies:

\begin{itemize}[itemsep = 4pt, topsep = 4pt]
    
    \item The representative-particle method uses a limited number of Lagrangian particles (e.g., $10^5$--$10^6$) to model the astrophysically large number of dust grains in real protoplanetary disks (e.g., $>10^{30}$) that cannot be handled directly by numerical models. Each {\it representative particle} statistically represents a group of {\it physical dust grains} with the same modeled dust properties. Its dust properties and statistical weight are updated through sampled collision events, while its position evolves according to the adopted dust-transport model. With a sufficient number of representative particles and sampled collision events, the evolution of the dust population can then be statistically reproduced through a Monte Carlo method (\citealt{zsom.A.2008.10,beutel.M.2023.02,beutel.M.2024.04}). One typical code using this method is \texttt{mcdust} \citep{vaikundaraman.V.2026.07}.
    
    \item The Eulerian mass-bin method discretizes the grain-mass distribution into multiple bins and models each of them as a distinct {\it pressureless dust fluid}. This method solves the Smoluchowski equation (\citealt{smoluchowski.MV.1916.01}), an integro-differential equation, by calculating collision rates and redistributing mass within and across these grain-size bins according to the prescribed collision outcomes. Examples include \texttt{DustPy} (\citealt{stammler.SM.2022.08}), \texttt{cuDisc} (\citealt{robinson.A.2024.04,robinson.A.2026.08}), and a new standalone solver \texttt{GRACE-DG} (\citealt{yang.J.2026.09}).
    
    \item The reduced two-population method approximates the dust population using only small- and large-grain components plus a characteristic maximum grain size. Instead of solving the full Smoluchowski equation, it uses simplified prescriptions for coagulation, fragmentation, and the partition of mass between the two populations. Examples include \texttt{two-pop-py} (\citealt{birnstiel.T.2012.03}), its successor \texttt{TriPoD} (\citealp{pfeil.T.2024.11,pfeil.T.2026.09}), and the open-source version \texttt{TriPoDPy} (\citealt{kaufmann.NL.2025}).

\end{itemize}

Apart from the recently published \texttt{GRACE-DG}, the other codes mentioned above are compared across a range of outputs by \citet{eriksson.LEJ.2026.07}, and we refer readers to that paper for a more detailed summary of the codes. Despite the availability of multiple dust evolution codes for planet formation studies, a few methodological gaps still persist:

\begin{itemize}[itemsep = 4pt, topsep = 4pt]

    \item Three-dimensional (3D) model: \texttt{cuDisc} and \texttt{mcdust} provide two-dimensional (2D) radial--vertical models, while the other open-source codes are one-dimensional (1D), except for \texttt{GRACE-DG}, which is a local (0D) solver. A 3D version of the \texttt{TriPoD}--\texttt{Athena++} framework is mentioned in \citet{eriksson.LEJ.2026.07}, but it relies on the reduced two-population approximation and, to our knowledge, has not yet been publicly released. Therefore, no open-source code resolves the full collisional evolution of the grain-size distribution together with the vertical distribution of dust, non-axisymmetric structures, and their combined effects on dust collisions.
    
    \item Dust momentum treatment: Among the codes above, only \texttt{cuDisc} and the \texttt{TriPoD}--\texttt{Athena++} framework provide dust transport with independently evolved momentum (\citealt{huang.P.2022.09,pfeil.T.2026.09}). All others use reduced dust transport with prescribed velocities. These prescriptions restrict dust motion to the assumed drift and settling regimes, potentially misrepresenting dust concentration and collision velocities where the underlying force-balance assumptions fail.
    
    \item General methodological issues: Eulerian mass-bin methods face a multiplicative increase in the total number of bins when additional dust properties are resolved as independent dimensions. In addition, insufficient mass-bin resolution can introduce numerical diffusion that artificially accelerates grain growth (\citealt{ohtsuki.K.1990.01,drazkowska.J.2014.07}). Representative-particle methods can retain sampled dust trajectories and properties, but finite sampling may underresolve low-mass regions (\citealt{zsom.A.2008.10}). Both approaches therefore face a trade-off between the fidelity of coupled collisional and dynamical evolution and the feasibility of numerical implementation.

\end{itemize}

Motivated by these limitations, we introduce \texttt{GameDev}, a $\boldsymbol{G}$PU-$\boldsymbol{a}$ccelerated $\boldsymbol{m}$od$\boldsymbol{e}$l for $\boldsymbol{d}$ust $\boldsymbol{ev}$olution that runs on graphics processing units (GPUs)\footnote{The code supports and has been tested on NVIDIA devices with CUDA 12.1 and AMD devices with ROCm 7.2. The code is publicly available at \href{https://github.com/bijiaqing/GameDev/}{https://github.com/bijiaqing/GameDev/}}. \texttt{GameDev} supports dust dynamics in 3D through both Lagrangian particle and Eulerian fluid formulations, with independently evolved particle velocities and fluid momentum, respectively. The particle formulation couples dust dynamics to Monte Carlo dust collisions, allowing dust grains undergoing collisional evolution to move without prescribing equilibrium drift or settling velocities.

The paper is organized as follows. Section~\ref{sec:particle} describes the particle model of \texttt{GameDev}, including gas prescriptions, dust transport, diffusion, collision, and numerical stepping. Section~\ref{sec:tests} presents numerical tests of the particle model, including particle trajectories, dust diffusion, grain coagulation against analytical solutions, followed by a comparison with other dust evolution codes. Section~\ref{sec:discussion} demonstrates an application of \texttt{GameDev} to the photospheric instability, discusses the computational performance, and outlines future extensions. We conclude in Sect.~\ref{sec:conclusions}. The appendices provide a description of the Eulerian dust fluid model, and a list of symbols.

\section{The particle model and physical prescriptions} \label{sec:particle}

The particle model describes dust evolution around a central star in a prescribed gas disk. The dynamical forcing of dust is provided by stellar gravity and aerodynamic drag, with optional radiation pressure and Poynting--Robertson drag. Dust diffusion is treated through stochastic spatial displacements, while grain growth and fragmentation are treated through sampled collision events. The current grain model assumes spherical grains with a fixed internal density, so that grain mass is uniquely determined by size. We specify the particle nature of the model here to distinguish it from the standalone Eulerian dust fluid model also provided by \texttt{GameDev} and described in Appendix~\ref{sec:fluid}.

The physical dust grain population is represented by an ensemble of $N_{\rm p}$ Lagrangian representative particles (hereafter, particles). Each particle $i$ carries a position vector $\boldsymbol{x}_i$, a velocity vector $\boldsymbol{v}_i$, a grain size $s_i$, and the number $N_i$ of represented physical dust grains. Taking $s_i$ as the diameter, the grain mass and particle mass of each particle are
\begin{equation}
    m_i = \pi\rho_\circ s_i^3 / 6
\end{equation}
and
\begin{equation}
    M_i = N_i m_i,
\end{equation}
where $\rho_\circ$ is the constant internal density of grains. 

\subsection{Geometry and dimensional reductions} \label{sec:geom}

\texttt{GameDev} is formulated in spherical coordinates $\left\{r, \theta, \phi\right\}$ centered on the star, where $r$ is the distance from the star, $\theta$ is the polar angle, and $\phi$ is the azimuthal angle. In this paper, we also use $\left\{R, Z\right\}$ to denote the cylindrical radius and height, respectively. 

The particle model supports four spatial configurations: full 3D, 2D radial--polar, 2D radial--azimuthal, and 1D radial. Axisymmetric models retain azimuthal angular momentum, while vertically integrated models constrain particles to the midplane, set polar motion to zero, and represent dust throughout the vertical column. These models evolve the surface density $\Sigma_{\rm d}$ instead of the volume density $\rho_{\rm d}$ and require a prescription for the unresolved vertical structure when calculating collision rates. Hereafter, we use $\varrho_{\rm d}$ and $\varrho_{\rm g}$ to denote the dust and gas densities: $\Sigma_{\rm d}$ and $\Sigma_{\rm g}$ in vertically integrated models, and $\rho_{\rm d}$ and $\rho_{\rm g}$ otherwise.

For configurations with a resolved azimuthal dimension, the domain may cover either the full azimuth or a periodic sector. A periodic sector assumes that the modeled distribution repeats outside the simulated interval. 

\subsection{The prescribed gas background} \label{sec:gas}

Explicit gas hydrodynamics is not included in \texttt{GameDev}. The gas background is either prescribed analytically or supplied through imported density and velocity fields. In the latter case, the gas fields are reloaded at each simulation output time, and the fields between successive loadings are linearly interpolated in time. In either case, feedback from dust onto the gas is not included.

When adopting the analytical prescription, the gas has a locally isothermal equation of state and is vertically isothermal. We assume an axisymmetric gas disk with power-law profiles for the gas surface density and aspect ratio, written as\footnote{We use the subscript 0 to denote evaluations in the disk midplane at the reference radius $R = R_0$, unless otherwise specified.}
\begin{equation}
    \Sigma_{\rm g}(R) = \Sigma_{\rm g0}\left(R/R_0\right)^p
\end{equation}
and
\begin{equation}
    h_{\rm g}(R) = h_{\rm g0}\left(R/R_0\right)^{(q + 1)/2}
\end{equation}
where $p$ and $q$ are the constant power-law indices of the gas surface density and temperature, respectively. The gas scale height and sound speed are given by
\begin{equation}
    H_{\rm g}(R) = h_{\rm g}R
\end{equation}
and
\begin{equation}
    c_{\rm s}(R) = H_{\rm g}\Omega_{\rm K},
\end{equation}
where 
\begin{equation}
    \Omega_{\rm K}(R) = \sqrt{GM_\star / R^3}
\end{equation}
is the Keplerian angular velocity, $G$ is the gravitational constant, and $M_\star$ is the stellar mass.

In configurations with a resolved vertical dimension, the gas volume density follows the hydrostatic-equilibrium solution
\begin{equation}
    \rho_{\rm g}\left(R, Z\right)
        = \frac{\Sigma_{\rm g}}{\sqrt{2\pi}H_{\rm g}}\times\exp\left(-\frac{1 - R/r}{h_{\rm g}^2}\right),
\end{equation}
which reduces to a Gaussian profile near the midplane. The analytical azimuthal velocity of the gas is given by
\begin{equation}
    v_{{\rm g},\phi}\left(R, Z\right) = \sqrt{1 - 2\eta}\,\Omega_{\rm K}R,
\end{equation}
where
\begin{equation}
    \eta\left(R, Z\right) = -\frac{1}{2}\left[
        \left(p + \frac{q}{2} - \frac{3}{2}\right)h_{\rm g}^2 + q\left(1 - \frac{R}{r}\right)
    \right]
\end{equation}
is the radial pressure-gradient term. The radial and vertical velocities of the gas in cylindrical coordinates are set to zero unless a viscous flow of the gas is included. The kinematic viscosity $\nu$ of the gas is either held constant globally or calculated using an otherwise constant $\alpha$ viscosity parameter (\citealt{shakura.NI.1973.01})
\begin{equation} 
    \nu(R) = \alpha c_{\rm s}H_{\rm g}.
\end{equation}
The viscous flow is imposed as a radial velocity of the gas following \citet{kanagawa.KD.2017.08}
\begin{equation} \label{eq:v_visc}
    v_{\rm visc}\left(R, Z\right) = -\frac{\nu}{R}\left[\,
        3\frac{\partial\ln\left(\nu\varrho_{\rm g}R^{1/2}\right)}{\partial\ln R}
        - q\frac{\partial\ln\left(\varrho_{\rm g}Z\right)}{\partial\ln Z}
    \,\right].
\end{equation}
The last term of Eq.~(\ref{eq:v_visc}) is dropped in vertically integrated configurations, so the equation naturally reduces to the conventional form in \citet{lynden-bell.D.1974.09}.

\subsection{Dust dynamics and advection} \label{sec:dyn}

The equation of motion for the dust is given by
\begin{equation}
    \frac{d\boldsymbol{v}}{dt}\left(\boldsymbol{x}, \boldsymbol{v}, s\right)
        = -\frac{GM_\star}{r^2}\hat{\boldsymbol{r}} + \boldsymbol{f}_{\rm rad} + \boldsymbol{f}_{\rm aero},
\end{equation}
where $\boldsymbol{f}_{\rm rad}$ and $\boldsymbol{f}_{\rm aero}$ are the specific radiation and aerodynamic forces, respectively.

\subsubsection{Aerodynamic drag} \label{sec:drag}

Aerodynamic drag relaxes the dust velocity toward the local velocity of the gas on the stopping timescale $t_{\rm s}$. We assume that dust grains are in the Epstein regime (\citealt{epstein.PS.1924.06}), so the drag term is given by
\begin{equation}
    \boldsymbol{f}_{\rm aero}\left(\boldsymbol{x}, \boldsymbol{v}, s\right)
        = \left(\boldsymbol{v}_{\rm g} - \boldsymbol{v}\right) /\,t_{\rm s},
\end{equation}
and the stopping time is prescribed as
\begin{equation}
    t_{\rm s}\left(\boldsymbol{x}, s\right)
        = \frac{s}{s_0}\left(\frac{\rho_{\rm g}}{\rho_{\rm g0}}\frac{c_{\rm s}}{c_{\rm s0}}\right)^{-1}
        \frac{{\rm St}_0}{\Omega_{\rm K0}}.
\end{equation}
Here, $s_0$ is a spatially invariant reference grain size, and the dimensionless Stokes number
\begin{equation}
    {\rm St}\left(\boldsymbol{x}, s\right) = \Omega_{\rm K}t_{\rm s}
\end{equation}
is the parameter we use to adjust the aerodynamic coupling between gas and dust in our models. For the analytical gas background, we have
\begin{equation}
    {\rm St}\left(R, Z, s\right)
        = {\rm St}_0\frac{s}{s_0}\left(\frac{R}{R_0}\right)^{-p}
        \times\exp\left(\frac{1 - R/r}{h_{\rm g}^2}\right).
\end{equation}
For the imported gas background, the Stokes number scales with the normalized density of the supplied gas via
\begin{equation}
    {\rm St}\left(\boldsymbol{x}, s\right)
        = {\rm St}_0\frac{s}{s_0}
        \left(\frac{\rho_{\rm g}}{\rho_{\rm g0}}\right)^{-1}
        \left(\frac{R}{R_0}\right)^{-(q + 3)/2}
\end{equation}
when the vertical dimension is resolved, and 
\begin{equation}
    {\rm St}\left(\boldsymbol{x}, s\right)
        = {\rm St}_0\frac{s}{s_0}
        \left(\frac{\Sigma_{\rm g}}{\Sigma_{\rm g0}}\right)^{-1}
\end{equation}
in vertically integrated models. Gas fields are supplied on numerical grids and interpolated to each particle's position using multilinear spatial interpolation.

\subsubsection{Radiation forces} \label{sec:rad}

\texttt{GameDev} supports an optional implementation of radiation pressure and Poynting--Robertson (P--R) drag
\begin{equation}
    \boldsymbol{f}_{\rm rad}\left(\boldsymbol{x}, \boldsymbol{v}, s\right)
        = \boldsymbol{f}_{\rm RP} + \boldsymbol{f}_{\rm PR}.
\end{equation}
Both forces scale with the $\beta$ parameter, which describes the strength of radiation pressure relative to gravity
\begin{equation} \label{eq:beta_def}
    \beta(s) = \frac{\kappa L_\star}{4\pi c\,GM_\star},
\end{equation}
where $\kappa$ is the dust opacity, $L_\star$ is the stellar luminosity, and $c$ is the speed of light. We prescribe the grain-size dependence of $\kappa$ following \citet{burns.JA.1979.10}
\begin{equation} \label{eq:kappa}
    \kappa(s) = \kappa_0\left(s/s_0\right)^{-1},
\end{equation}
therefore, we also have
\begin{equation} \label{eq:beta}
    \beta(s) = \beta_0\left(s/s_0\right)^{-1}.
\end{equation}
The radiation pressure is then given by 
\begin{equation}
    \boldsymbol{f}_{\rm RP}\left(\boldsymbol{x}, s\right)
        = \beta{\rm e}^{-\tau}\frac{GM_\star}{r^2}\hat{\boldsymbol{r}},
\end{equation}
where $\tau$ is the local optical depth measured in the radial direction from the inner boundary assuming no additional interior extinction, and is given by
\begin{equation} \label{eq:tau_3d}
    \tau\left(\boldsymbol{x}\right) = \int_{r_{\rm min}}^{r}\int_{0}^{\infty}
        \kappa\left(s^\prime\right)
        \frac{\partial\rho_{\rm d}}{\partial s^\prime}\left(r^\prime, s^\prime\right)ds^\prime dr^\prime
\end{equation}
when the vertical dimension is resolved\footnote{Despite the relation in Eqs.~(\ref{eq:beta_def}), (\ref{eq:kappa}), and (\ref{eq:beta}), $\kappa_0$ and $\beta_0$ can be specified independently in \texttt{GameDev} to decouple the prescriptions for the radiation forces from those for the optical depth.}. In vertically integrated models, we evaluate $\tau$ in the midplane while assuming that all dust grains share the same scale height, equal to that of the gas, regardless of grain size, which gives
\begin{equation} \label{eq:tau_2d}
    \begin{aligned}
        \tau\left(\boldsymbol{x}\right) &= \\[4pt]
        \int_{R_{\rm min}}^{R} & \frac{1}{\sqrt{2\pi}H_{\rm g}\left(R^\prime\right)}
            \int_{0}^{\infty}\kappa\left(s^\prime\right)
            \frac{\partial\Sigma_{\rm d}}{\partial s^\prime}\left(R^\prime, s^\prime\right)
            ds^\prime dR^\prime.
    \end{aligned}
\end{equation}
In practice, particles are first interpolated onto the numerical grids to obtain $\tau$. Similar to the treatment of the imported gas fields described in Sect.~\ref{sec:drag}, the $\tau$ field is then interpolated back to each particle's position during force evaluations.

The P--R drag introduces velocity-dependent damping effects. To first order in $\boldsymbol{v}/c$, it can be written as
\begin{equation}
    \boldsymbol{f}_{\rm PR}\left(\boldsymbol{x}, \boldsymbol{v}, s\right)
        = -\gamma_{\rm PR}\left(
            v_\theta\hat{\boldsymbol{\theta}} + 2v_r\hat{\boldsymbol{r}} + v_\phi\hat{\boldsymbol{\phi}}
        \right),
\end{equation}
where 
\begin{equation}
    \gamma_{\rm PR}\left(\boldsymbol{x}, s\right)
        = \frac{\beta{\rm e}^{-\tau}}{c}\frac{GM_\star}{r^2}.
\end{equation}
Unlike radiation pressure, which is treated as an explicit forcing term during the numerical integration, the P--R damping terms are included in combination with the aerodynamic drag terms (see Sect.~\ref{sec:ssa}). 

To avoid introducing transient effects at the beginning of a simulation, radiation forces are usually multiplied by a taper factor $\xi^2\left(3 - 2\xi\right)$, where
\begin{equation}
    \xi(t) = \min\Big[\max\left(t / t_\beta,\,0\right),\,1\,\Big],
\end{equation}
such that radiation forces are gradually turned on over the timescale $t_\beta$.

\subsubsection{The staggered semi-analytic trajectory integration} \label{sec:ssa}

We use the staggered semi-analytic method introduced in \citet{fung.J.2019.10} to advance the positions and dynamical variables of the particles in spherical coordinates. This method is second-order accurate when the stopping time $t_{\rm s}$ is resolved by the numerical timestep $\Delta t$, and it approaches a leapfrog scheme in the drag-free limit. In the stiff regime where $\Delta t \gg t_{\rm s}$, its analytical drag treatment and staggered force evaluation accurately recover the terminal drift without resolving the rapid velocity relaxation.

For computational efficiency, we evolve the integration variables
\begin{equation}
    \boldsymbol{y} = \left(v_r, \ell_\theta, \ell_\phi\right)
\end{equation}
which use the specific angular momenta instead of the linear momenta in the polar and azimuthal directions, where
\begin{equation}
    \ell_\theta = rv_\theta
\end{equation}
and 
\begin{equation}
    \ell_\phi = v_\phi r\sin\theta.
\end{equation}
The equation of motion can then be written as
\begin{equation}
    \frac{d\boldsymbol{y}}{dt}
        = -\mathbf{\Gamma}\boldsymbol{y} + \frac{\boldsymbol{y}_{\rm g}}{t_{\rm s}} + \boldsymbol{f},
\end{equation}
where the diagonal damping-rate matrix is
\begin{equation}
    \mathbf{\Gamma} = {\rm diag}\left(\gamma_r, \gamma_\theta, \gamma_\phi\right),
\end{equation}
with
\begin{equation}
    \gamma_r = t_{\rm s}^{-1} + 2\gamma_{\rm PR}
\end{equation}
and
\begin{equation}
    \gamma_\theta = \gamma_\phi = t_{\rm s}^{-1} + \gamma_{\rm PR}.
\end{equation}
The remaining specific force terms are
\begin{align}
    & f_r = -\left(1 - \beta{\rm e}^{-\tau}\right)\frac{GM_\star}{r^2}
        + \frac{\ell_\theta^2}{r^3} + \frac{\ell_\phi^2}{r^3\sin^2\theta}, \\[4pt]
    & f_\theta = \frac{\ell_\phi^2\cos\theta}{r^2\sin^3\theta},
\end{align}
and
\begin{equation}
    f_\phi = 0.
\end{equation}

\subsection{Dust diffusion} \label{sec:diff}

In addition to the deterministic motion described above, particles undergo stochastic spatial redistribution to represent dust diffusion. We prescribe a dust diffusion tensor that is diagonal in the local cylindrical basis
\begin{equation}
    \mathbf{D} = {\rm diag}\left(D_R, D_Z, D_\phi\right),
\end{equation}
where the diffusion coefficient of imperfectly coupled dust grains in each direction is approximated as
\begin{equation}
    D = \frac{\nu}{{\rm Sc}\left(1 + {\rm St}^2\right)}.
\end{equation}
Here, ${\rm Sc}$ is the Schmidt number in each direction and is the primary parameter through which we prescribe the diffusion level independently of the direction and strength of viscous flows of the gas.

We provide two implementations of dust diffusion, with their Fickian target equations written as
\begin{equation} \label{eq:diff_conc}
    \left.\frac{\partial\varrho_{\rm d}}{\partial t}\right|_{\rm diffusion}
        = \nabla\cdot\left[
            \varrho_{\rm g}\mathbf{D}\nabla\left(\frac{\varrho_{\rm d}}{\varrho_{\rm g}}\right)
        \right]
\end{equation}
and
\begin{equation} \label{eq:diff_dens}
    \left.\frac{\partial\varrho_{\rm d}}{\partial t}\right|_{\rm diffusion}
        = \nabla\cdot\left(\mathbf{D}\nabla\varrho_{\rm d}\right).
\end{equation}
Equation~(\ref{eq:diff_conc}) acts on the dust-to-gas ratio and is typically used for turbulent mixing (e.g., \citealt{ciesla.FJ.2010.11,charnoz.S.2011.08,zsom.A.2011.10,desch.SJ.2017.05}). We also provide Eq.~(\ref{eq:diff_dens}) as an alternative. It acts on the dust density and can provide an effective description of redistribution driven by particle interactions, such as collisional spreading in particulate disks (\citealt{quillen.AC.2007.05}). For vertically integrated models, the vertical terms are omitted. The azimuthal terms are likewise omitted in axisymmetric models.

We reproduce Eqs.~(\ref{eq:diff_conc}) and (\ref{eq:diff_dens}) statistically using an Euler--Maruyama update (\citealt{kloeden.PE.1992}). Over a diffusion timestep $\Delta t_{\rm diff}$, particle displacements in cylindrical coordinates are calculated as
\begin{equation} \label{eq:step_R}
    \begin{aligned}
        \Delta R = {} & \sqrt{2D_R\Delta t_{\rm diff}}\zeta_R \\[4pt]
        & + \left(
            \frac{D_R}{R} + \frac{\partial D_R}{\partial R}
            + \chi D_R\frac{\partial\ln\varrho_{\rm g}}{\partial R}
        \right)\Delta t_{\rm diff},
    \end{aligned}
\end{equation}
\begin{equation}
    \begin{aligned}
        \Delta Z = {} & \sqrt{2D_Z\Delta t_{\rm diff}}\zeta_Z \\[4pt]
        & + \left(
            \frac{\partial D_Z}{\partial Z} + \chi D_Z\frac{\partial\ln\varrho_{\rm g}}{\partial Z}
        \right)\Delta t_{\rm diff},
    \end{aligned}
\end{equation}
and 
\begin{equation} \label{eq:step_phi}
    \begin{aligned}
        \Delta\phi = {} & \sqrt{2D_\phi\Delta t_{\rm diff}}\frac{\zeta_\phi}{R} \\[4pt]
        & + \left(
            \frac{\partial D_\phi}{\partial\phi}
            + \chi D_\phi\frac{\partial\ln\varrho_{\rm g}}{\partial\phi}
        \right)\frac{\Delta t_{\rm diff}}{R^2},
    \end{aligned}
\end{equation}
where $\chi = 1$ for Eq.~(\ref{eq:diff_conc}), $\chi = 0$ for Eq.~(\ref{eq:diff_dens}), and the $\zeta$ values are independent random numbers drawn from a normal distribution with zero mean and unit variance $\mathcal{N}\left(0, 1\right)$. The displacements are then converted to the coordinates used to store particle positions and applied to the particles. This redistribution process does not change the magnitude or direction of the dust velocity in Cartesian coordinates. We recalculate the stored velocity components to account for the change in the local coordinate basis at the new position.

\subsection{Dust collisions}

Dust transport determines where grains encounter one another and the local populations available for collisions. We model collisional evolution by sampling encounters between particles and updating their dust properties according to prescribed collision outcomes. Changes in grain size then modify the aerodynamic coupling and radiation forces governing subsequent particle motion.

\subsubsection{Collision rate} \label{sec:rate}

We estimate the local collisional environment of each particle $i$ using its $N_{\rm K}$ nearest particles within a user-specified maximum search-scale, including the query particle at zero distance\footnote{Including the query particle allows collisions with physical dust grains belonging to its own dust population, which should not be neglected in the statistical sampling (\citealt{zsom.A.2008.10,beutel.M.2023.02}). For $N_i\gg1$, we approximate the number of available partners, $N_i-1$, by $N_i$.}. The nearest neighbors are identified using either the k-dimensional tree (KD-tree) method adapted from the \texttt{cudaKDTree} library (\citealt{bentley.JL.1975.09,wald.I.2022.10}), or the adaptive spatial hierarchy method based on the Morton order (\citealt{morton.GM.1966}). Both search methods are available because their relative performance depends on the model configuration.

The neighborhood size $N_{\rm K}$ controls the balance between statistical sampling and spatial resolution. Larger neighborhoods improve sampling but average over a broader space, potentially smoothing local variations. The convergence of statistically sampled collision outcomes toward the analytical solution for varying $N_{\rm K}$ is examined in Sect.~\ref{sec:test_coag} and Appendix~\ref{sec:survey}. The importance of adequate local sampling and spatial resolution in particle collision models was also demonstrated by \citet{drazkowska.J.2013.08}.

Let $\mathcal{J}_i$ denote the retained neighbors of particle $i$, and let $d_i$ denote the distance to the farthest neighbor in $\mathcal{J}_i$. In 3D models, the sampling region is a sphere with volume
\begin{equation}
    V_i = 4\pi d_i^3 / 3.
\end{equation}
Each neighbor $j$ represents $N_j = M_j / m_j$ physical dust grains, which we treat as uniformly distributed within this region when estimating the collision rate of $i$. Its contribution to the local grain-number density is therefore
\begin{equation}
    n_{ij} = N_j / V_i.
\end{equation}
For spherical grains with diameters $s_i$ and $s_j$, the collision kernel is the geometrical cross section multiplied by the relative speed,
\begin{equation} \label{eq:kernel}
    K_{ij} = \pi\left(s_i + s_j\right)^2\Delta v_{ij} / 4,
\end{equation}
where $\Delta v_{ij}$ is the relative speed specified in Sect.~\ref{sec:outcome}. The collision rate of a physical dust grain represented by particle $i$ with the population represented by neighbor $j$ is then
\begin{equation} \label{eq:rate}
    \lambda_{ij} = n_{ij}K_{ij} = N_jK_{ij} / V_i.
\end{equation}
Summing over the neighbors gives the collision rate of particle $i$
\begin{equation}
    \lambda_i = \sum_{j\in\mathcal{J}_i}\lambda_{ij}.
\end{equation}

The sampling measure $V_i$ is adjusted to the model geometry. When the vertical dimension is resolved, $V_i$ is a volume. In 2D radial--polar models, neighbors are identified in the meridional plane, and the sampled area is converted to a toroidal volume by multiplying it by the local circumference $2\pi R_i$. In vertically integrated models, neighbors are identified in the midplane, and $V_i$ is instead the sampled area. In 2D radial--azimuthal models, it is a disk of radius $d_i$. In 1D radial models, it is an annulus that spans $\left[R_i - d_i, R_i + d_i\right]$. The vertical length scale is then supplied by assuming that the dust grains of population $i$ are vertically distributed following the normal probability density with zero mean and variance $H_i^2$
\begin{equation}
    \mathcal{N}\left(Z; 0, H_i^2\right) =
        \frac{1}{\sqrt{2\pi}H_i}\times\exp\left(-\frac{Z^2}{2H_i^2}\right),
\end{equation}
where $H_i = H_{\rm g}\left(R_i\right)$. The grain-number density of population $j$ at height $Z$ is then
\begin{equation}
    \begin{aligned}
        n_{ij}\left(Z\right) & = \frac{N_j}{V_i}\mathcal{N}\left(Z; 0, H_j^2\right) \\[4pt]
            & = \frac{N_j}{\sqrt{2\pi}H_jV_i}\times\exp\left(-\frac{Z^2}{2H_j^2}\right).
    \end{aligned}
\end{equation}
Treating $K_{ij}$ as independent of height, we average the collision rate over the vertical distribution of population $i$
\begin{equation}
    \lambda_{ij} = K_{ij}\int_{-\infty}^{\infty}
        \mathcal{N}\left(Z; 0, H_i^2\right)n_{ij}\left(Z\right)dZ,
\end{equation}
which finally gives
\begin{equation}
    \lambda_{ij} = \frac{N_jK_{ij}}{V_i\sqrt{2\pi\left(H_i^2 + H_j^2\right)}}.
\end{equation}
Near the domain boundaries, we reduce the sampling measure to account for the portion outside the domain, approximating each nearby boundary as locally flat. Particles across a periodic azimuthal boundary are also considered as candidates in the neighbor search.

\subsubsection{Collision speed and prescribed outcomes} \label{sec:outcome}

The collision speed combines prescribed differential motions with contributions from Brownian motion and unresolved turbulence
\begin{equation} \label{eq:dv}
    \Delta v_{ij}^2 =
        \Delta v_R^2 + \Delta v_Z^2 + \Delta v_\phi^2 + \Delta v_{\rm B}^2 + \Delta v_{\rm t}^2.
\end{equation}
The relative-motion terms are given by
\begin{align}
    & \Delta v_R = 2\eta R\Omega_{\rm K}\left(
        \frac{{\rm St}_i}{1 + {\rm St}_i^2} - \frac{{\rm St}_j}{1 + {\rm St}_j^2}
    \right), \\[8pt]
    & \Delta v_Z = Z\Omega_{\rm K}\Big[
        \min\left({\rm St}_i, 0.5\right) - \min\left({\rm St}_j, 0.5\right)
    \Big],
\end{align}
and
\begin{equation}
    \Delta v_\phi = \eta R\Omega_{\rm K}\left(
        \frac{1}{1 + {\rm St}_i^2} - \frac{1}{1 + {\rm St}_j^2}
    \right).
\end{equation}
We adopt these prescribed differential velocities here because the resolved velocities of neighbors separated by up to $d_i$ also include the background shear across the neighborhood. Dust momentum therefore affects collisions through the spatial distribution of dust rather than through the collision speeds. The Brownian contribution is written as
\begin{equation}
    \Delta v_{\rm B} = \min\left[\,
        c_{\rm s}, \sqrt{\frac{8k_{\rm B}T}{\pi}\left(\frac{1}{m_i} + \frac{1}{m_j}\right)}
    \,\right],
\end{equation}
where $k_{\rm B}$ is the Boltzmann constant and $T$ is the gas temperature. All the variables entering Eq.~(\ref{eq:dv}) are evaluated at the query particle's position. We include Brownian motion only in calculations with explicit physical units. For the turbulent contribution, we adopt the six-branch piecewise expressions of \citet{ormel.CW.2007.05}, adapted from the \texttt{DustPy} implementation. We note that the Brownian and turbulent terms specify unresolved contributions to the collision speed and do not apply additional velocity kicks to the particles. Similarly, the stochastic spatial displacements introduced in Sect.~\ref{sec:diff} do not contribute to collision speeds.

We distinguish collision outcomes among sticking, erosion, and fragmentation using the prescribed fragmentation speed $v_{\rm frag}$ and the grain-mass ratio 
\begin{equation}
    \psi = m_j / m_i.
\end{equation}
Collisions with $\Delta v_{ij} < v_{\rm frag}$ result in the sticking of physical dust grains, with the new grain mass given by
\begin{equation}
    m_i^\prime\Big|_{\rm sticking} = m_i + m_j.
\end{equation}
Given the fixed internal density of grains adopted here, its diameter then becomes
\begin{equation}
    s_i^\prime\Big|_{\rm sticking} = \sqrt[3]{s_i^3 + s_j^3}.
\end{equation}
Collisions with $\Delta v_{ij} \geq v_{\rm frag}$ and $\psi \leq 0.1$ result in erosion, following the prescription of \citet{birnstiel.T.2011.01} as implemented in \texttt{mcdust}. The probability that the particle follows the remnant or the debris depends on the grain-mass ratio, such that
\begin{equation}
    m_i^\prime\Big|_{\rm erosion} =
    \begin{cases}
        m_i - m_j, & \text{with probability } 1 - \psi, \\[4pt]
        m_j, & \text{with probability } \psi.
    \end{cases}
\end{equation}
Notably, to improve numerical performance, small sticking increments and erosive decrements are grouped in the collisional calculations following \citet{ormel.CW.2008.09} and \citet{zsom.A.2008.10}, similar to \texttt{mcdust}. The grouping factor is given by
\begin{equation}
    N_{\rm coll} =
    \begin{cases}
        \left\lfloor 10^{-4}/\psi\right\rfloor, & \psi \leq 10^{-6}, \\[4pt]
        1, & \text{otherwise},
    \end{cases}
\end{equation}
and the grouped sticking gives
\begin{equation}
    m_i^\prime\Big|_{\rm sticking} = m_i + N_{\rm coll}m_j
\end{equation}
at a reduced effective collision rate
\begin{equation}
    \lambda_{ij}^*\Big|_{\rm sticking} = N_{\rm coll}^{-1}\lambda_{ij}.
\end{equation}
Here, the effective collision rate $\lambda_{ij}^*$ describes both grouped and ungrouped collision events, with $\lambda_{ij}^* = \lambda_{ij}$ for $N_{\rm coll} = 1$. Grouped erosion retains separate prescriptions for the remnant and the debris with
\begin{equation}
    m_i^\prime\Big|_{\rm erosion} =
    \begin{cases}
        m_i - N_{\rm coll}m_j,
            & \lambda_{ij,{\rm rem}}^* = N_{\rm coll}^{-1}\left(1 - \psi\right)\lambda_{ij}, \\[4pt]
        m_j, & \lambda_{ij,{\rm deb}}^* = \psi\lambda_{ij}
    \end{cases}
\end{equation}
at an effective collision rate
\begin{equation}
    \lambda_{ij}^*\Big|_{\rm erosion} = \lambda_{ij,{\rm rem}}^* + \lambda_{ij,{\rm deb}}^*.
\end{equation}
Other high-speed collisions lead to fragmentation. The new diameter assigned to the particle is sampled as
\begin{equation} \label{eq:frag}
    s_i^\prime\Big|_{\rm fragment} = \Big[
        \sqrt{s_{\rm min}} + U\left(\sqrt{s_i} - \sqrt{s_{\rm min}}\right)
    \Big]^2,
\end{equation}
where $U \sim \mathcal{U}\left(0, 1\right)$ is a random number drawn from a uniform distribution and $s_{\rm min}$ is the prescribed minimum grain size. Equation~(\ref{eq:frag}) samples the fragment diameter from the mass-weighted distribution
\begin{equation}
    dM_{\rm frag} / ds \propto s^{-1/2},
\end{equation}
which corresponds to the conventional grain-number distribution in \citet{mathis.JS.1977.10}
\begin{equation}
    dN_{\rm frag} / ds \propto s^{-7/2}.
\end{equation}
After every collision, the number of physical dust grains represented by the particle is updated while its particle mass $M_i$ is preserved
\begin{equation}
    N_i^\prime = M_i / m_i^\prime.
\end{equation}

\subsubsection{Monte Carlo sampling and particle updates} \label{sec:mc}

During each collision timestep $\Delta t_{\rm coll}$, particles are advanced through a sequence of randomly sampled collision events. However, the outcome of a collision event involving one particle is not immediately broadcast to others. The collisional evolution follows the workflow below.

The spatial domain is partitioned into multiple {\it spatial groups}. For each local spatial group, the globally shared collision timestep $\Delta t_{\rm coll}$ is divided into one or more adaptive {\it frozen-neighbor intervals} $\Delta t_{\rm fn}$. Unless otherwise specified, the quantities introduced below refer to a single spatial group.

At the beginning of each frozen-neighbor interval $\Delta t_{\rm fn}$, a particle caches the broadcast properties of its $N_{\rm K}$ nearest neighbors. It then evolves independently against this frozen neighborhood. A particle's properties are updated after every collision event, with multiple events allowed within the same interval. However, the particles in a local spatial group broadcast their updated properties and refresh the cached properties of their neighbors only after completing their local interval. This approximation permits the parallel evolution of all particles, but a neighborhood-refresh query may access properties broadcast at earlier times because different spatial groups may have different frozen-neighbor intervals.

The length of the frozen-neighbor interval is controlled by the expected change in the grain size of the local dust population. Within each local spatial group, particles are assigned into a set $\mathcal{B}$ of adaptive logarithmic grain-size bins, which we call {\it controller bins}\footnote{Adjacent sparsely populated grain-size bins are merged to improve statistical sampling.}. We evaluate the first and second moments of the grain-size change for each controller bin $b \in \mathcal{B}$ at the beginning of each frozen-neighbor interval as
\begin{equation}
    A_b = \frac{
        \displaystyle\sum_{i\in b}M_i\sum_o\lambda_{io}^*\left\langle J_{io}\right\rangle
    }{
        \displaystyle\sum_{i\in b}M_i
    },
\end{equation}
and
\begin{equation}
    B_b = \frac{
        \displaystyle\sum_{i\in b}M_i\sum_o\lambda_{io}^*\left\langle J_{io}^2\right\rangle
    }{
        \displaystyle\sum_{i\in b}M_i
    },
\end{equation}
where
\begin{equation}
    J_{io} = \left|\,\ln\left(s_i^\prime / s_i\right)\,\right|
\end{equation}
is a dimensionless parameter of the grain-size change for a sampled collision channel $o$ acting on particle $i$, and $\lambda_{io}^*$ is its sampling rate. The quantity $\left\langle J_{io}\right\rangle$ denotes the expectation over the random collision outcome, with $\left\langle J_{io}\right\rangle = J_{io}$ for deterministic outcomes. This expectation is introduced because the fragmentation outcome depends on the random draw $U$ (see Eq.~(\ref{eq:frag})), giving
\begin{equation}
    \left\langle J_{io}\right\rangle\Big|_{\rm fragment} = \int_{0}^{1}J_{io}(U)\,dU.
\end{equation}

The first moment $A_b$ measures how quickly grain sizes change, and the second moment $B_b$ measures the corresponding fluctuation induced by random collision events. Using a prescribed dimensionless tolerance $\varepsilon$ to constrain the speed of grain-size changes and their fluctuations before refreshing the frozen neighborhood, we require any candidate frozen-neighbor interval to satisfy
\begin{equation}
    A_b\Delta t_{\rm fn} \leq \varepsilon C
\end{equation}
and
\begin{equation}
    B_b\Delta t_{\rm fn} \leq \varepsilon^2C^2,
\end{equation}
where $0.25 \leq C \leq 1$ is an adaptive safety factor shared by all controller bins $b \in \mathcal{B}$. The requested frozen-neighbor interval $\Delta t_{\rm fn,req}$ is then constrained by
\begin{equation}
    \Delta t_{\rm fn,req} = \min\left[\,
        \Delta t_{\rm remain},\,
        \Delta t_{\rm fn,max},\,
        \min_{b\in\mathcal{B}}\frac{\varepsilon C}{A_b},\,
        \min_{b\in\mathcal{B}}\frac{\varepsilon^2C^2}{B_b}
    \,\right],
\end{equation}
where $\Delta t_{\rm remain}$ is the remaining time to complete the current collision timestep and $\Delta t_{\rm fn,max}$ is the prescribed maximum frozen-neighbor interval. The safety factor $C$ is initialized to 1 and inherited between adjacent frozen-neighbor intervals. It is adjusted by post-interval audits, which monitor the collision frequency, the grain-size changes, and the redistribution of mass between controller bins. While post-interval audits adjust the length of subsequent intervals, completed collision events are not rejected or repeated.

Using the requested interval $\Delta t_{\rm fn,req}$ as a reference, the executed frozen-neighbor interval $\Delta t_{\rm fn,exe}$ is obtained by repeatedly halving the collision timestep $\Delta t_{\rm coll}$. To limit differences in how frequently interacting groups refresh their neighbor properties, the ratios of their intervals are restricted to at most 16. However, timestep changes respect already scheduled intervals, so a requested refinement may be delayed.

Within each frozen-neighbor interval, a particle $i$ samples a waiting time until its next collision event following \citet{gillespie.DT.1975.10}
\begin{equation}
    \Delta t_{\rm w} = -\ln U_i / \lambda_i^*,
\end{equation}
where $U_i \sim \mathcal{U}\left(0, 1\right)$ is, again, a random number drawn from a uniform distribution and 
\begin{equation}
    \lambda_i^* = \sum_{j\in\mathcal{J}_i}\lambda_{ij}^*.
\end{equation}
If this waiting time exceeds the remaining interval, no further collision occurs. Otherwise, the particle advances its clock by $\Delta t_{\rm w}$, undergoes an event, and updates its own dust properties and effective collision rate $\lambda_i^*$. For each collision event involving particle $i$, a partner population $j$ is selected with probability $\lambda_{ij}^*/\lambda_i^*$. Then the collision event follows the outcome prescriptions described in Sect.~\ref{sec:outcome}. 

Although implicit in the workflow, each collision event updates only the corresponding particle $i$. Any particle $j$ supplying the partner properties evolves through its own independently sampled events. Despite this asymmetry, a sufficiently large particle number $N_{\rm p}$ and sufficiently short frozen-neighbor intervals controlled by $\varepsilon$ should allow the ensemble to statistically capture the evolution of the dust population without updating both participants of the same event. This method is tested in Sect.~\ref{sec:test_coag} and Appendix~\ref{sec:survey}.

\subsection{Numerical stepping} \label{sec:step}

In \texttt{GameDev}, we combine dust dynamics, diffusion, and collisional evolution through a symmetric sequence of operator updates. For each numerical timestep $\Delta t$, we first advance collisions and diffusion by half a step each, then integrate particle trajectories over the full step, and repeat diffusion and collisions in reverse order
\begin{equation*}
    \mathcal{C}\left(\frac{\Delta t}{2}\right)
    \mathcal{D}\left(\frac{\Delta t}{2}\right)
    \mathcal{T}\left(\vphantom{\frac{\Delta t}{2}}\Delta t\right)
    \mathcal{D}\left(\frac{\Delta t}{2}\right)
    \mathcal{C}\left(\frac{\Delta t}{2}\right),
\end{equation*}
where $\mathcal{C}$, $\mathcal{D}$, and $\mathcal{T}$ are operators denoting collision, diffusion, and deterministic transport, respectively, such that $\Delta t_{\rm coll} = \Delta t_{\rm diff} = \Delta t/2$. This symmetric operator splitting is adopted to reduce splitting errors (\citealp{strang.G.1968.09}). Collision operators are placed outside the diffusion operators so that the cached nearest-neighbor lists can be reused between the final collision operator of one timestep and the first collision operator of the next.

All active particles share a common transport timestep, selected from the following constraints
\begin{equation}
    \Delta t = \min\left[\,
        \Delta t_{\rm max},\,
        \Delta t_{\rm out},\,
        \min_i\Delta t_i
    \,\right],
\end{equation}
where $\Delta t_{\rm max}$ is the prescribed maximum timestep, $\Delta t_{\rm out}$ is the remaining time before the next output, and $\Delta t_i$ is the most restrictive local timestep for particle $i$, determined by orbital motion and by deterministic and diffusive displacements relative to the local grid spacing.

\section{Numerical tests} \label{sec:tests}

In this section, we present numerical tests of \texttt{GameDev}'s particle model and validate its accuracy in particle trajectories, dust diffusion, and collisional coagulation against analytical solutions. Finally, we reproduce the 2D advection--diffusion--collision model of \citet{eriksson.LEJ.2026.07} and compare the numerical results with those of other existing models. All \texttt{GameDev} test sets are available on GitHub.

\subsection{Particle trajectory tests} \label{sec:test_orbit}

\begin{figure}[t]
\centering
\includegraphics[width = \columnwidth]{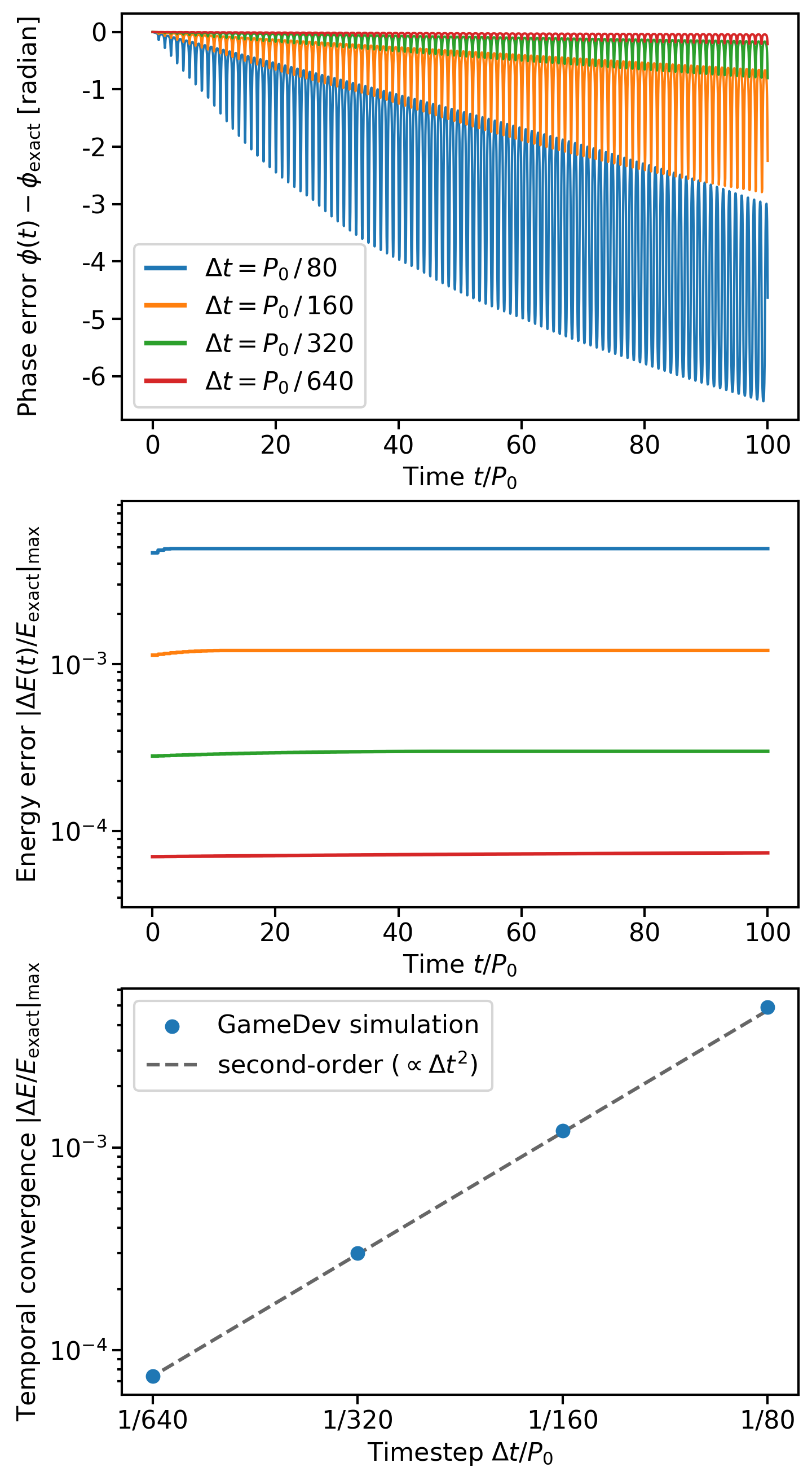}
\caption{
    Eccentric-orbit test for a particle initialized on a drag-free eccentric Keplerian orbit with $e = 0.5$ and integrated with four fixed timesteps. Top: phase error. Middle: running maximum of the relative energy error. Bottom: maximum relative energy error over $100P_0$ as a function of timestep, with the dashed line indicating second-order convergence.
\label{fig:orbit}}
\end{figure}

We tested the accuracy of a particle's trajectory by simulating its motion on an eccentric Keplerian orbit in a gravity-only, drag-free disk model. Eccentric Keplerian orbits provide exact periodic references with strongly varying velocity and acceleration, allowing repeated passages through the pericenter to expose accumulated phase errors and artificial orbital precession alongside errors in conserved quantities.

Using code units with $GM_\star = R_0 = 1$, we initialized a single particle in a 2D radial--azimuthal disk model with
\begin{equation*}
    \left(R,\,\phi,\,v_r,\,\ell_\phi\right)
        = \left(1/2,\,0,\,0,\,\sqrt{3}/2\right).
\end{equation*}
This initialization places the particle at the pericenter of an eccentric Keplerian orbit with eccentricity $e = 0.5$, semi-major axis $R_0$, orbital period $P_0 = 2\pi$, and the exact specific orbital energy $E_{\rm exact} = -0.5$. The exact true anomaly is then given by
\begin{equation}
    \phi_{\rm exact}(t) = 
        2\tan^{-1}\left[\,
            \sqrt{\frac{1 + e}{1 - e}}\tan\frac{u}{2}
        \,\right],
\end{equation}
where $u(t)$ is the eccentric anomaly that satisfies
\begin{equation}
    u - e\sin u = t.
\end{equation}

We integrated the particle's trajectory for $100 P_0$ with four different fixed dynamical timesteps
\begin{equation*}
    \Delta t = \left\{
        \frac{P_0}{80},\,\frac{P_0}{160},\,\frac{P_0}{320},\,\frac{P_0}{640}
    \right\}.
\end{equation*}
The phase and energy errors are shown in Fig.~\ref{fig:orbit}. While the numerical trajectory develops a systematic phase lag, both the accumulated phase error and the energy error decrease as the timestep is reduced. The maximum relative energy error scales approximately as $\Delta t^2$, consistent with second-order temporal convergence, and falls below $10^{-4}$ for $\Delta t=P_0/640$.

\subsection{Dust diffusion tests} \label{sec:test_diff}

\begin{figure*}[t]
\sidecaption
\includegraphics[width = 12cm]{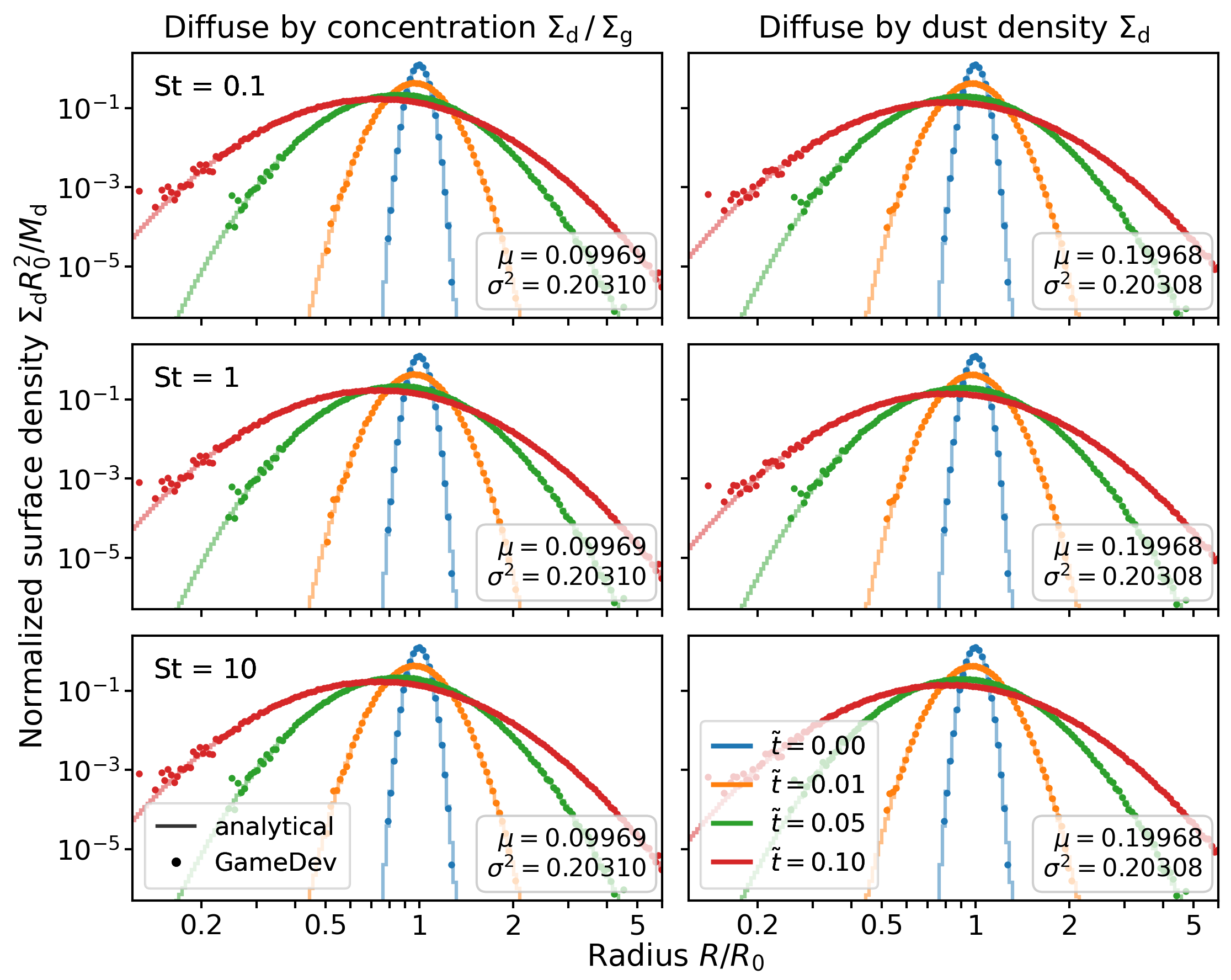}
\caption{
    Dust-diffusion tests with concentration diffusion (Eq.~(\ref{eq:diff_conc}); left) and density diffusion (Eq.~(\ref{eq:diff_dens}); right) for ${\rm St} = 0.1$, 1, and 10 (top to bottom). Dots show the simulated surface densities at $\tilde{t} = 0$, 0.01, 0.05, and 0.1, and lines show the corresponding analytical solutions. Each panel lists the measured mean logarithmic radius $\mu$ and variance $\sigma^2$ at $\tilde{t} = 0.1$.
\label{fig:diff}}
\end{figure*}

We tested the accuracy of dust diffusion by simulating the expansion of a dust ring due to stochastic diffusive displacements of particles in a 1D radial disk model without external forcing such as gravity and aerodynamic drag. The gas disk was prescribed with $h_{\rm g0} = 0.05$, $p = -1$, $q = 0.5$, and $\alpha = 0.04$. 

We initialized $2^{20}$ particles with zero velocity and assigned each particle $i$ a random position according to
\begin{equation}
    \ln\left(R_i/R_0\right) \Big|_{t = 0}
        \sim \mathcal{N}\Big(\,\mu(0),\,\sigma^2(0)\Big),
\end{equation}
where $\mu(0) = 0$ and $\sigma(0) = 0.05$ are the initial mean and standard deviation of the logarithmic radius, respectively. Assuming that all particles carry the same particle mass, the same spatially constant Stokes number, and ${\rm Sc} = 1$, the diffusion coefficient of dust can be written as
\begin{equation} \label{eq:ring_D}
    D(R) = D_0\left(R/R_0\right)^2,
\end{equation}
where
\begin{equation}
    D_0 = 10^{-4}/(1 + {\rm St}^2)
\end{equation}
is a numerical value in code units. The initial surface density of dust is
\begin{equation} \label{eq:ring_init}
    \Sigma_{\rm d}\left(R, t = 0\right)
        = \frac{M_{\rm d}}{\sqrt{8\pi^3}R^2\sigma(0)}
        \exp\left\{-\frac{1}{2}\left[\frac{\ln\left(R/R_0\right)}{\sigma(0)}\right]^2\right\},
\end{equation}
where $M_{\rm d}$ is the total dust mass. We tested both diffusion prescriptions described in Eqs.~(\ref{eq:diff_conc}) and (\ref{eq:diff_dens}) using
\begin{equation} \label{eq:ring_pde}
    \frac{\partial\Sigma_{\rm d}}{\partial t}\left(R, t\right)
        = \frac{1}{R}\frac{\partial}{\partial R}\left[
            RD\left(
                \frac{\partial\Sigma_{\rm d}}{\partial R}
                - \chi\Sigma_{\rm d}\frac{\partial\ln\Sigma_{\rm g}}{\partial R}
            \right)
        \right],
\end{equation}
where $\chi = 1$ for Eq.~(\ref{eq:diff_conc}) and $\chi = 0$ for Eq.~(\ref{eq:diff_dens}), as in Eqs.~(\ref{eq:step_R})--(\ref{eq:step_phi}). Combining Eqs.~(\ref{eq:ring_init}) and (\ref{eq:ring_pde}), the analytical solution of the time-dependent density profile is
\begin{equation}
    \Sigma_{\rm d}\left(R, t\right)
        = \frac{M_{\rm d}}{\sqrt{8\pi^3}R^2\sigma(t)}
        \exp\left\{-\frac{1}{2}\left[\frac{\ln\left(R/R_0\right) - \mu(t)}{\sigma(t)}\right]^2\right\},
\end{equation}
where
\begin{equation}
    \mu(t)
        = \Big\langle\ln\left(R_i / R_0\right)\Big\rangle_i
        = \left(2 + \chi p\right)D_0 t / R_0^2
\end{equation}
tracks the displacement of the dust distribution, and
\begin{equation}
    \sigma^2(t)
        = \Big\langle\Big[\,\ln\left(R_i / R_0\right) - \mu\,\Big]^2\Big\rangle_i
        = \sigma^2(0) + 2D_0 t / R_0^2
\end{equation}
measures its broadening. For each diffusion prescription, we established three models with different Stokes numbers
\begin{equation*}
    {\rm St} = \left\{0.1,\,1,\,10\right\}.
\end{equation*}
The Stokes number enters the dust diffusivity in Eq.~(\ref{eq:ring_D}) and therefore affects the dimensionless time
\begin{equation}
    \tilde{t} = t/t_0 = D_0 t / R_0^2,
\end{equation}
where 
\begin{equation}
    t_0 = R_0^2 / D_0
\end{equation}
is the reference time of the diffusion test.
We simulated each model to $\tilde{t} = 0.1$, so that the resulting structures could be compared at the same evolutionary stage, and the results are shown in Fig.~\ref{fig:diff}.

Both diffusion prescriptions in \texttt{GameDev} accurately reproduce the diffusive evolution of dust rings compared with their corresponding analytical solutions. At $\tilde{t} = 0.1$, the measured mean logarithmic radii differ from the analytical values by 0.31\% for concentration diffusion and 0.16\% for density diffusion. The measured variances agree with the analytical value of 0.2025 to approximately 0.3\%. Panels in the same column coincide because the models used the same dimensionless times, dimensionless timesteps, and random-number sequences. While a larger scatter of the profiles in the low-density regions is consistent with finite particle sampling, the scatter patterns are also expected to be similar across the two columns because of the shared random-number sequence.

\subsection{Grain coagulation tests} \label{sec:test_coag}

\begin{figure}[t]
\centering
\includegraphics[width = \columnwidth]{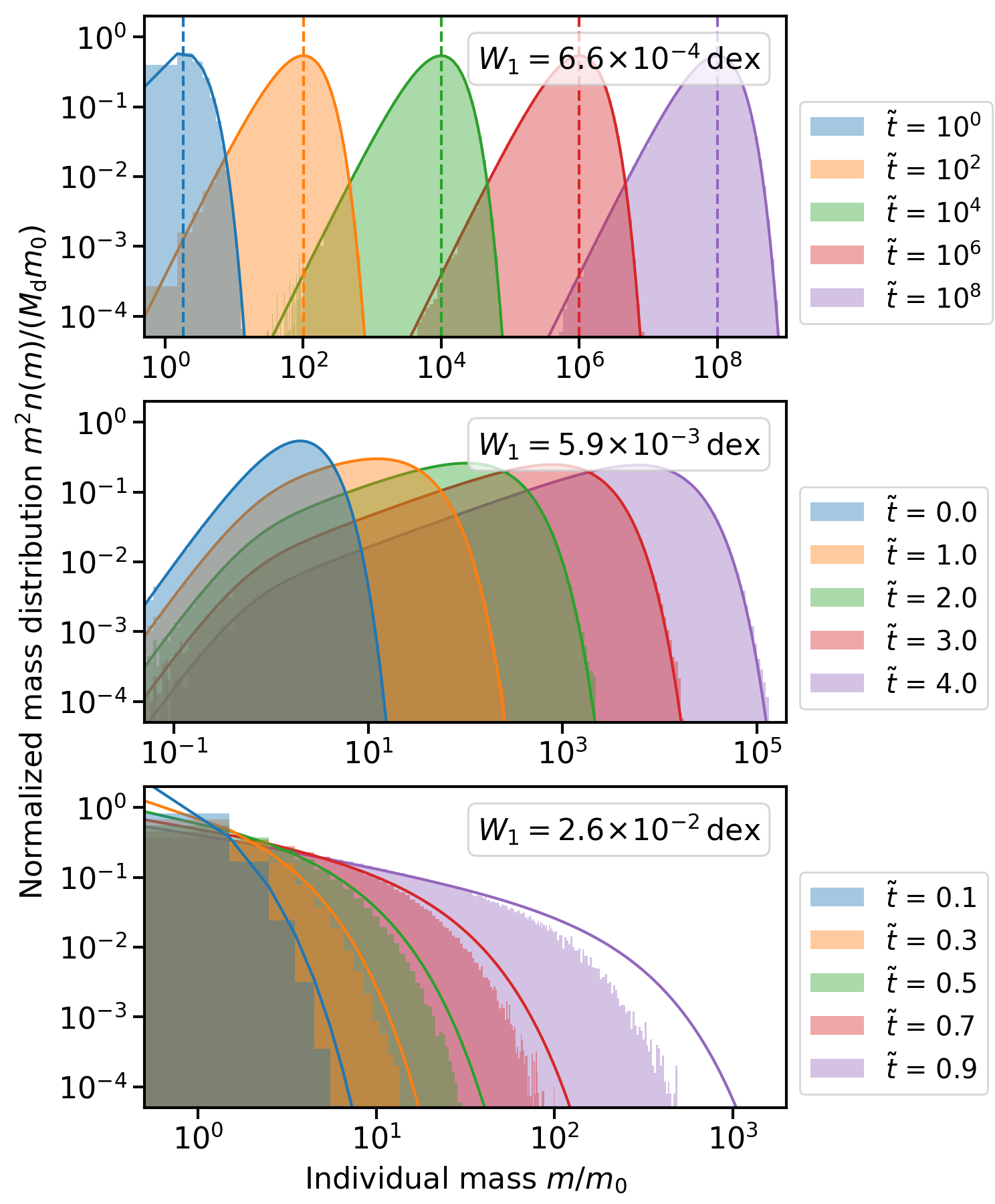}
\caption{
    Normalized grain-mass distributions in the constant (top), linear (middle), and product (bottom) kernel tests with $N_{\rm K} = 256$ and $\varepsilon = 0.08$. Shaded histograms show the simulation results and solid curves show the analytical solutions at the dimensionless times $\tilde{t}$ given in the legends. Vertical dashed lines mark the analytical peak masses $m_{\rm peak}$ in the constant-kernel test given by Eq.~(\ref{eq:m_peak}). Each panel lists the error indicator, the Wasserstein distance $W_1$, at the final snapshot, which is averaged over ten repetitions (Appendix~\ref{sec:survey}).
\label{fig:coag}}
\end{figure}

We tested the accuracy of dust collisions and coagulation by simulating the time evolution of the grain-size distribution of a group of particles undergoing collisions without spatial motion. We prescribed coagulation-only outcomes with three idealized collision kernels, namely the constant, linear, and product kernels, for which the Smoluchowski equation has closed-form analytical solutions provided in Appendix~\ref{sec:smol}.

We initialized $N_{\rm p} = 10^6$ particles with equal particle masses
\begin{equation}
    M_i = M_{\rm d} / N_{\rm p}
\end{equation}
inside a 2D annulus with $0.5 < R/R_0 < 1.5$, $-\pi < \phi < \pi$, and $\theta = \pi/2$, where $M_{\rm d}$ is the total dust mass. Particles were close to uniformly distributed in $R$ and $\phi$, with small independent displacements from a strictly uniform distribution to prevent excessive equal-distance ties in neighbor searches. For all coagulation models, we define
\begin{equation}
    \lambda_0 = \frac{N_{\rm p}m_0}{N_{\rm K}M_{\rm d}}\frac{1}{t_0},
\end{equation}
where $m_0$ is the unit mass and $t_0$ is the unit time, so that the dimensionless time can be written as
\begin{equation}
    \tilde{t} = \frac{N_{\rm K}M_{\rm d}}{N_{\rm p}m_0}\lambda_0 t = \frac{t}{t_0},
\end{equation}
and collision rates were calculated using an adapted form of Eq.~(\ref{eq:rate}) with a prescribed dimensionless measure of the neighborhood $V_i = 1$
\begin{equation}
    \begin{aligned}
        \lambda_{ij}\left(m_i, m_j\right)
        & = \lambda_0\frac{N_j}{V_i}K_{ij}\left(m_i, m_j\right) \\[4pt]
        & = \frac{1}{t_0}\frac{1}{N_{\rm K}}\frac{m_0}{m_j}K_{ij}\left(m_i, m_j\right).
    \end{aligned}
\end{equation}

As mentioned in Sects.~\ref{sec:rate} and \ref{sec:mc}, $N_{\rm K}$ controls the quality of statistical sampling of collision events with the nearest-neighbor approach while $\varepsilon$ controls the refresh intervals in the frozen-neighbor scheme. Both are newly introduced components of our collision model, and their parameters jointly affect the accuracy of the grain-size evolution and the computational cost. To identify parameters that provide a suitable balance between accuracy and computational performance, we explored combinations of
\begin{equation*}
    N_{\rm K} = \left\{16,\,32,\,64,\,128,\,256\right\}
\end{equation*}
and 
\begin{equation*}
    \varepsilon = \left\{0.01,\,0.02,\,0.04,\,0.08,\,0.16\right\}.
\end{equation*}
A summary of the coagulation evolution compared with the analytical solutions is provided in Figs.~\ref{fig:coag_const}, \ref{fig:coag_linear}, and \ref{fig:coag_prod}.

We measured the elapsed time for each model to complete its final snapshot and quantified the discrepancy between the simulation results and the analytical solutions using the dimensionless Wasserstein distance (\citealt{vallender.SS.1974.09}). Both metrics are described in more detail in Appendix~\ref{sec:survey}, with an accuracy-versus-performance plot provided in Fig.~\ref{fig:coag_cost}.

We find that $N_{\rm K} = 256$, the largest value tested, is strongly preferred to achieve good agreement with the analytical solutions, and $\varepsilon = 0.08$ generally provides good accuracy across all three kernel tests. A comparison of the simulation results from this model with the analytical solutions for the three kernel tests is provided in Fig.~\ref{fig:coag}. Meanwhile, given that physical collision kernels
\begin{equation*}
    K_{ij} \propto \left(s_i + s_j\right)^2\Delta v_{ij}
\end{equation*}
scale approximately linearly with grain mass when drift, settling, or turbulence dominates the collision speed, we suggest that $\varepsilon = 0.04$ and $\varepsilon = 0.16$ remain suitable choices when greater emphasis is placed on accuracy or computational performance, respectively.

\subsection{Benchmark comparison} \label{sec:test_e26}

\begin{figure*}[t]
\centering
\includegraphics[width = \textwidth]{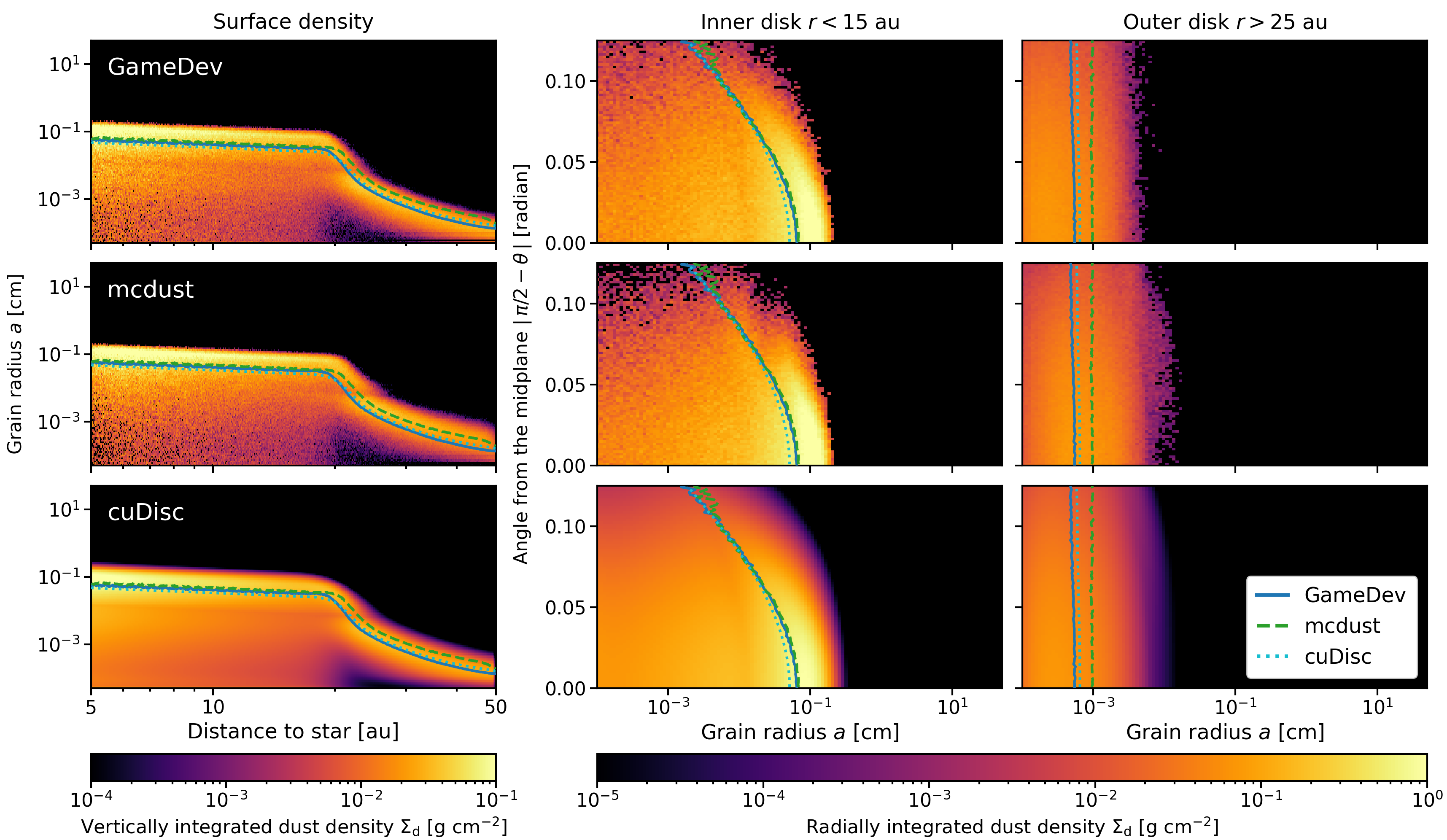}
\caption{
    Partial reproduction of Fig.~14 of \citet{eriksson.LEJ.2026.07} for the strong-turbulence model with $\alpha = 10^{-3}$ at $t = 11{,}100$~yr, showing \texttt{GameDev} (top), \texttt{mcdust} (middle), and \texttt{cuDisc} (bottom, at $t = 11{,}134$~yr). Left: vertically integrated surface density of dust per grain-size bin as a function of the distance to the star. Middle and right: radially integrated surface density of dust per grain-size bin as a function of the angle from the midplane in the inner ($r < 15$~au) and outer ($r > 25$~au) disk. Lines show the mass-weighted mean radius of grains in each model and are overplotted in all rows for comparison.
\label{fig:e26_1}}
\end{figure*}

\begin{figure*}[t]
\centering
\includegraphics[width = \textwidth]{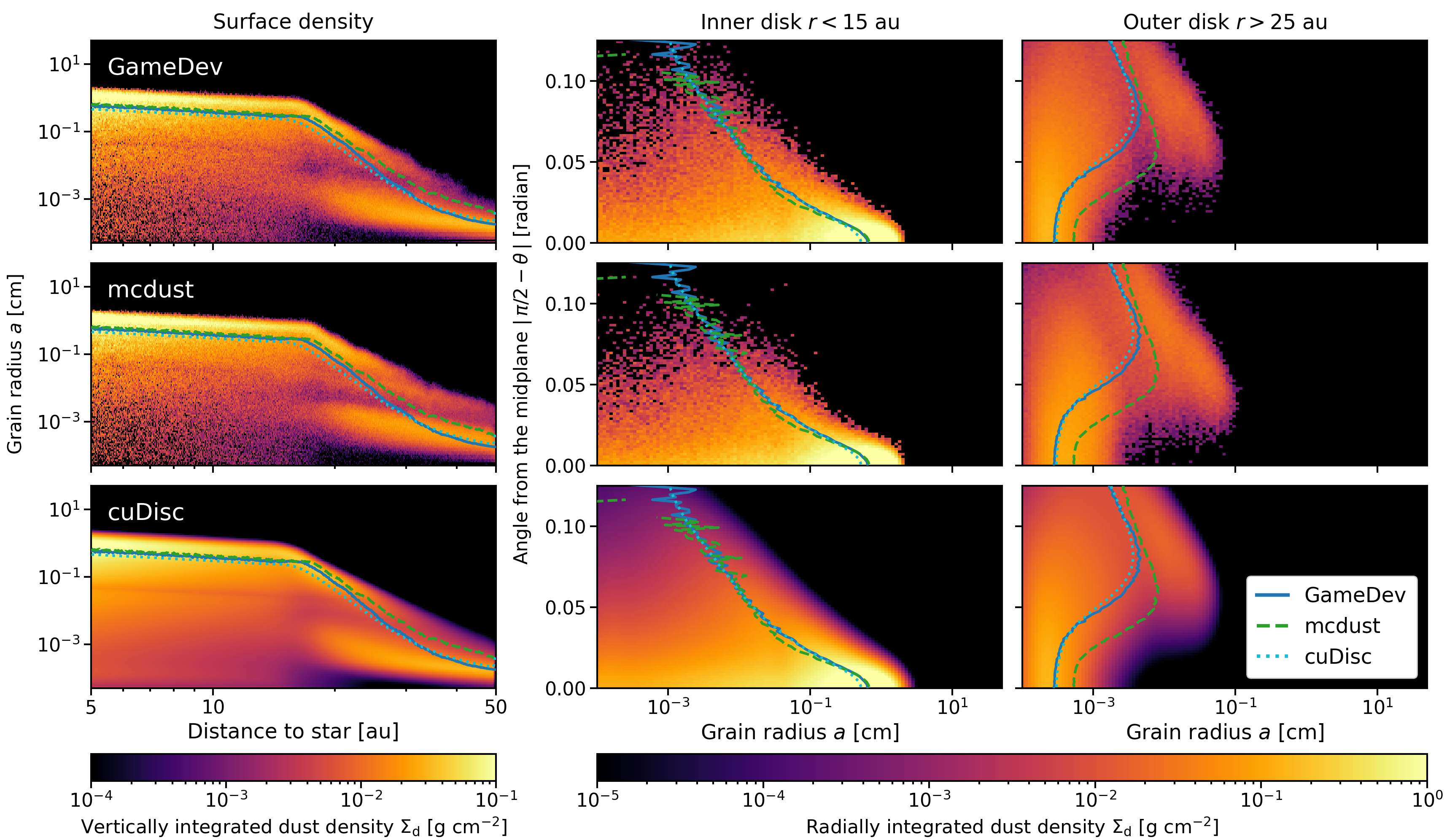}
\caption{
    Same as Fig.~\ref{fig:e26_1} but for the weak-turbulence model with $\alpha = 10^{-4}$ at $t = 23{,}800$~yr ($t = 23{,}809$~yr for \texttt{cuDisc}), reproducing Fig.~16 of \citet{eriksson.LEJ.2026.07}.
\label{fig:e26_2}}
\end{figure*}

\begin{figure*}[t]
\centering
\includegraphics[width = \textwidth]{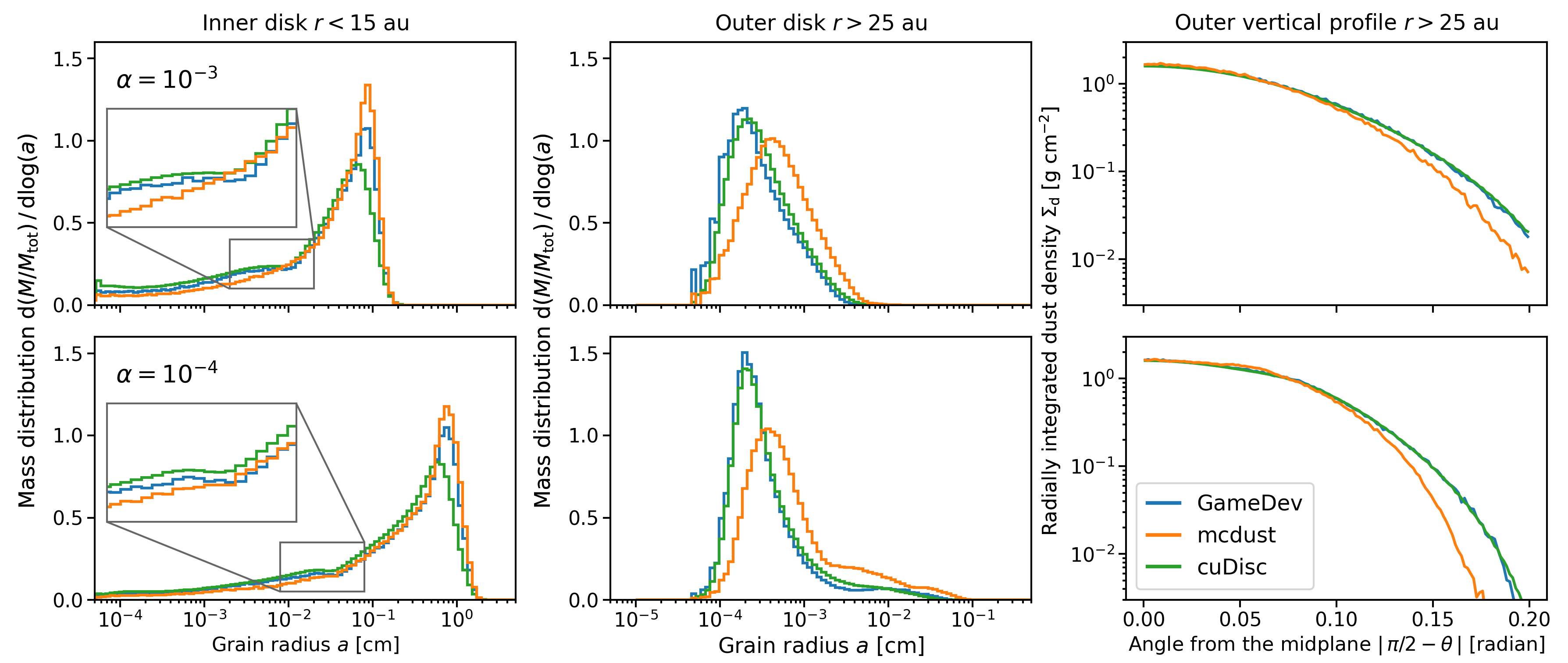}
\caption{
    Grain-mass distributions and vertical profiles of the \citet{eriksson.LEJ.2026.07} models with $\alpha = 10^{-3}$ at $t = 11{,}100$~yr (top) and $\alpha = 10^{-4}$ at $t = 23{,}800$~yr (bottom). Left and middle: mass-weighted distributions of grain radius in the inner ($r < 15$~au) and outer ($r > 25$~au) disk, with insets in the left panels magnifying the plateau feature. Right: radially integrated surface density of dust in the outer disk as a function of the angle from the midplane.
\label{fig:e26_3}}
\end{figure*}

While the measured accuracy and performance of the code, as well as the preferred choices of $N_{\rm K}$ and $\varepsilon$, may depend on the idealized coagulation tests, we further constructed more realistic models to provide more representative evidence. \citet{eriksson.LEJ.2026.07} (hereafter \citetalias{eriksson.LEJ.2026.07}) present a comparison of several dust evolution models. They consider a specific setup for sedimentation-driven coagulation, which is both scientifically relevant and numerically well-suited for further testing \citep{dullemond.CP.2005.05,tanaka.H.2005.05,zsom.A.2011.10,drazkowska.J.2013.08}. 

Following Sect.~5 in \citetalias{eriksson.LEJ.2026.07}, we built both the strong-turbulence model with $\alpha = 10^{-3}$ and the weak-turbulence model with $\alpha = 10^{-4}$ in \texttt{GameDev} with $N_{\rm p} = 2^{20}$, $N_{\rm K} = 256$, and $\varepsilon = 0.08$. We also performed matched runs with the Lagrangian particle model \texttt{mcdust}, which runs on central processing units (CPUs), with the same $N_{\rm p}$ and 256 particles per collision cell. Results from the two models were further compared with those from the GPU-based Eulerian fluid model \texttt{cuDisc} provided by \citetalias{eriksson.LEJ.2026.07}, although small discrepancies in snapshot times exist. Figures~\ref{fig:e26_1} and \ref{fig:e26_2} are partial reproductions of Figs.~14 and 16 in \citetalias{eriksson.LEJ.2026.07}, with the addition of the \texttt{GameDev} results. A summary of the comparison is provided in Fig.~\ref{fig:e26_3}.

Consistent with \citetalias{eriksson.LEJ.2026.07}, we find good overall agreement among the models tested, including \texttt{GameDev}. However, in quantitative comparisons, \texttt{GameDev} agrees more closely with \texttt{cuDisc} than with \texttt{mcdust}, despite sharing a Lagrangian particle formulation with \texttt{mcdust} rather than the Eulerian fluid formulation of \texttt{cuDisc}. This contrast is especially obvious in the outer disk, where \texttt{mcdust} deviates from the other two in both the vertically and radially integrated surface-density distributions in Figs.~\ref{fig:e26_1} and \ref{fig:e26_2}. 

To investigate the origin of the discrepancy among the three models, we compared the grain-mass distributions in the inner and outer disks in Fig.~\ref{fig:e26_3}. In the inner disk, the three models generally agree well. However, unlike the other two models, \texttt{mcdust} does not show the ``plateau'' feature around grain radius $a = s/2 \sim 0.01$~cm in the $\alpha = 10^{-3}$ model and shows only a weak feature around $a \sim 0.03$~cm in the $\alpha = 10^{-4}$ model. In the outer disk, the differences become more pronounced. \texttt{mcdust} shifts the dust population toward larger grain sizes and also produces stronger sedimentation than the other two models. More rapid grain growth could enhance sedimentation, while stronger sedimentation could in turn promote grain growth. Because grain growth and sedimentation are coupled through this feedback loop, these results alone do not clearly identify the origin of the discrepancy.

However, one notable methodological difference is the treatment of dust transport. \texttt{GameDev} and \texttt{cuDisc} use different dust formulations but both evolve dust momentum independently, whereas \texttt{mcdust} adopts prescribed dust velocities. Such prescribed velocities may provide a good approximation in the inner disk but may become less accurate in the outer disk, where the aerodynamic coupling between dust and gas becomes weaker. We therefore tentatively attribute at least part of the discrepancy among the three models to the treatment of dust transport. This comparison highlights the potential importance of independently evolving dust momentum when modeling the coupled evolution of grain growth and sedimentation.

\section{Discussion} \label{sec:discussion}

\subsection{Application example} \label{sec:psi}

\begin{figure}[t]
\centering
\includegraphics[width = \columnwidth]{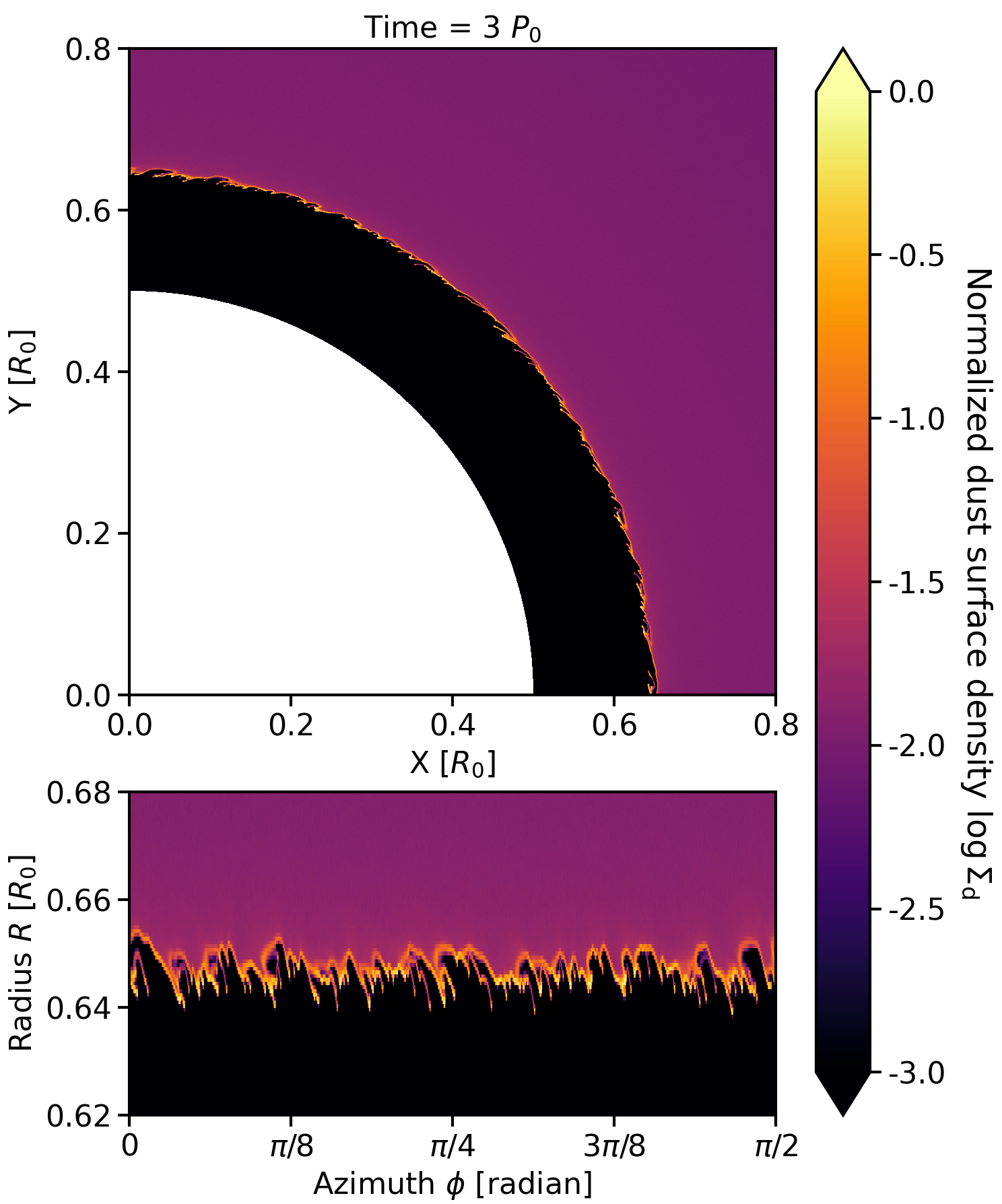}
\caption{
    Dust surface density in the particle model of the photospheric instability at $t = 3P_0$. Top: inner part of the simulated sector in Cartesian coordinates. Bottom: inner edge of the dust disk in the $R$--$\phi$ plane across the full sector.
\label{fig:psi}}
\end{figure}

The development of \texttt{GameDev} was primarily motivated by, and used for, studying the photospheric instability (PSI; \citealt{bi.J.2022.03}; Fung et al., submitted). The PSI is a radiation-driven instability that develops at the illuminated inner edge of dusty disks. There, radiation pressure sweeps dust outward into a sharp optically thin-to-thick transition, i.e., a dust wall. Small azimuthal density perturbations along the wall are then amplified: underdense segments are pushed outward faster by radiation pressure, sheared behind overdense segments, and radially compressed by shadowing. The instability breaks the axisymmetry of the disk and produces compact dust clumps. Radiation can then leak around these clumps to reach the previously shadowed material behind the wall. Consequently, the combination of radiation pressure, shadowing, and gas drag can drive a net outward recession of the dust disk edge, potentially clearing dust cavities of tens of au in gas-rich transitional disks. Moreover, the clumps may serve as sites for planetesimal formation. To probe dust evolution inside these clumps, it is therefore important to incorporate collisional evolution into the model of \citet{bi.J.2022.03}, or equivalently, to extend a dust evolution model to resolve the azimuthal direction.

Figure~\ref{fig:psi} shows the morphology of the dusty disk in a demonstration case where PSI is developing. We adopted a 2D radial--azimuthal model spanning $0.5 < R/R_0 < 1.5$ in a periodic sector $0 < \phi < \pi/2$, with code units $GM_\star = R_0 = 1$. The analytical gas disk had $p = -0.5$, $q = 0$, and $h_{\rm g0} = 0.05$. Dust of a single grain size was represented by $N_{\rm p} = 10^9$ particles with equal particle mass and evolved under stellar gravity, aerodynamic drag with ${\rm St}_0 = 10^{-3}$, and radiation pressure with $\beta_0 = 10$. Dust diffused following Eq.~(\ref{eq:diff_dens}) with $\alpha = 10^{-4}$ and ${\rm Sc} = 1$, while collisions and Poynting--Robertson drag were not included. The axisymmetric initial surface density of dust was a truncated power law smoothed by a Gaussian kernel
\begin{equation} \label{eq:psi_init}
    \begin{aligned}
        \Sigma_{\rm d}\left(R, t = 0\right) & = \\[4pt]
        \frac{\Sigma_{\rm d0}}{\sqrt{2\pi}w}
        & \int_{0.6R_0}^{1.4R_0}\left(\frac{R^\prime}{R_0}\right)^{p}
            \exp\left[-\frac{\left(R - R^\prime\right)^2}{2w^2}\right]dR^\prime,
    \end{aligned}
\end{equation}
where $w = 0.025R_0$. We prescribed $\kappa_0$ and $\Sigma_{\rm d0}$ such that the initial optical depth was $\tau \sim 6$ near the peak of the initial surface density, and $\tau$ was mapped on a $1024\times1536$ grid for radial and azimuthal resolutions. Particles were sampled from Eq.~(\ref{eq:psi_init}) with uniformly distributed azimuths and initialized with their drag-equilibrium drift velocities. No perturbation was imposed, so any non-axisymmetric structure grows from particle-sampling noise, whose relative amplitude was $\sim 3\%$ at the initial density peak. Follow-up studies on the PSI incorporating collisional dust evolution and on the PSI developing in the 3D photospheric layers of the disk will be presented in separate papers.

\subsection{Computational performance} \label{sec:perf}

We compared the elapsed computational time of the two Lagrangian particle models, \texttt{GameDev} and \texttt{mcdust}, for the benchmarks in Sect.~\ref{sec:test_e26}, where both models used $N_{\rm p} = 2^{20}$, \texttt{GameDev} used $N_{\rm K} = 256$ and $\varepsilon = 0.08$, and \texttt{mcdust} used 256 particles per collision cell. The results are summarized in Fig.~\ref{fig:perf}. Despite the differences in the simulation results, \texttt{GameDev} with the Morton-order neighbor search completes both benchmarks faster than \texttt{mcdust} when GPUs and CPUs of similar generations are compared. In this configuration, the Morton-order neighbor search is faster than the KD-tree on all three GPUs. We note that $\varepsilon$ may be increased further to improve performance without sacrificing much accuracy, given that physical collision kernels scale approximately linearly with grain mass (Sect.~\ref{sec:test_coag}).

Notably, each \texttt{GameDev} model currently uses only one GPU, which typically occupies between one-half and one-eighth of a high-performance computing node, whereas the 72- and 128-core \texttt{mcdust} models each occupied an entire node. This leaves greater room for scaling \texttt{GameDev} to larger models. With the planned extension to multi-GPU parallelization (Sect.~\ref{sec:gpu}), \texttt{GameDev} may also benefit from faster communication between devices within a node than between separate nodes, which may otherwise become a performance bottleneck when scaling CPU-based models further.

\subsection{Future extensions}
\subsubsection{Multi-GPU parallelization} \label{sec:gpu}

The current implementation of \texttt{GameDev} supports only one GPU for each model. Although the typical video random access memory (VRAM) capacity of a GPU should generally be sufficient for models commensurate with its computing power, dividing a single \texttt{GameDev} model into multiple partitions and distributing them across multiple GPUs remains a promising approach for further improving computational performance.

Collisional evolution is expected to benefit most from multi-GPU parallelization. This is because neighbor searching and repeated collision updates can account for a substantial fraction of the total computational cost, posing a major challenge when scaling the model up even when the VRAM capacity of individual GPUs is not a bottleneck. The Morton ordering implemented in \texttt{GameDev} for neighbor searching provides a natural basis for spatial domain decomposition by partitioning the ordered particles among GPUs. Distributed KD-trees offer another possible approach (\citealt{patwary.MMA.2016.05}). Efficient parallelization would therefore require suitable data exchange and load balancing, particularly when collision activity varies strongly across the disk. The relative scalability of these approaches remains to be established.

\subsubsection{N-body implementation}

A natural extension of \texttt{GameDev} is to couple its collisional evolution of dust grains to the growth and orbital dynamics of planetesimals and planetary embryos. Coagulation and fragmentation determine the particle sizes, radial drift, and duration of the pebble supply, thereby influencing planetary growth through pebble accretion. For example, fragmentation can promote core growth by maintaining efficiently accreted particle sizes and prolonging the delivery of solids (\citealt{drazkowska.J.2021.03}). Following this evolving reservoir could therefore reduce the reliance on prescribed pebble sizes and fluxes in population-synthesis calculations.

Such an extension could be achieved by retaining the statistical particle representation for pebbles and small dust grains while introducing a separate population of individually resolved planetesimals or planetary embryos. The mutual gravitational interactions among massive bodies could be evolved using an established integrator, such as \texttt{GENGA} or \texttt{SyMBA} (\citealt{duncan.MJ.1998.10,grimm.SL.2014.11,grimm.SL.2022.06,lau.TCH.2023.04}). While there are already precedents for coupling dust evolution to N-body dynamics, such as the \texttt{DustPy}--\texttt{SyMBAp} framework in \citet{lau.TCH.2024.08}, \texttt{GameDev} could retain the local spatial and velocity distributions of dust evolving through collisions alongside the dynamics of the accreting bodies.

With a consistent treatment of pebble accretion and the associated mass and momentum transfer, this coupling would allow planetesimals and planetary embryos to compete for a shared, evolving pebble reservoir. It could quantify how collisional processing changes accretion efficiencies and growth timescales, and when simplified pebble-accretion prescriptions remain adequate. Detailed calculations could then inform improved prescriptions for larger population-synthesis ensembles, even where directly resolving the coupled evolution in every system remains computationally expensive.

\section{Conclusions} \label{sec:conclusions}

We present \texttt{GameDev}, an accurate and efficient dust evolution model that is accelerated by graphics processing units (GPUs). \texttt{GameDev} couples independently evolved 3D dust dynamics with Monte Carlo collisional evolution. Its main features are summarized as follows:

\begin{itemize}[itemsep = 4pt, topsep = 4pt]

    \item The particle model of \texttt{GameDev} simulates dust grains using Lagrangian representative particles in a prescribed gas disk. It integrates particle trajectories under stellar gravity, aerodynamic drag, and optional radiation pressure and Poynting--Robertson drag using the staggered semi-analytic method, which converges at second order. The integration method remains accurate for both tightly and weakly coupled dust grains. Dust diffusion is implemented with the Euler--Maruyama method.
    
    \item Collision rates are estimated from the $N_{\rm K}$ nearest neighbors of each particle, found with either a KD-tree or a Morton-order search, so the sampling region adapts to the local dust distribution without a grid. Collisions result in sticking, erosion, or fragmentation depending on the collision speed and the mass ratio of the colliding grains. Each particle samples collision events against a neighborhood that remains frozen over adaptive intervals. The interval length is controlled by a tolerance on the expected grain-size change and its fluctuation, allowing all particles to evolve their collisions in parallel.

    \item \texttt{GameDev} agrees well with both \texttt{mcdust} and \texttt{cuDisc} in the benchmark of sedimentation-driven coagulation, but shows closer agreement with \texttt{cuDisc}. This suggests that independently evolved dust momentum is important for modeling the coupled evolution of grain growth and sedimentation. Idealized coagulation tests identify the combination of $N_{\rm K} = 256$ and $\varepsilon = 0.08$ as a suitable balance between accuracy and computational cost.
        
    \item \texttt{GameDev} supports both NVIDIA and AMD GPUs. With the Morton-order neighbor search, \texttt{GameDev} on one GPU completes the sedimentation-driven coagulation benchmark faster than \texttt{mcdust} on a full CPU node of a similar generation, while occupying only a fraction of a GPU node.
    
    \item The particle model of \texttt{GameDev} supports 3D, 2D radial--polar, 2D radial--azimuthal, and 1D radial configurations, with the collision sampling measure adjusted consistently to each geometry.

    \item A standalone Eulerian pressureless-fluid model complements the particle model where extremely weak perturbations must be resolved (see Appendix~\ref{sec:fluid}).
    
    \item The source code of \texttt{GameDev}, as well as the setup for all models and tests presented in this paper, is available on GitHub.

\end{itemize}

\begin{acknowledgements}
    We thank Hongping Deng, Ruobing Dong, Joanna Drążkowska, Kees Dullemond, Mario Flock, Jeffrey Fung, Ya-Ping Li, Min-Kai Lin, Beibei Liu, and many others for helpful discussions over the past few years since the project was initiated. This work is supported by the Deutsche Forschungsgemeinschaft (DFG) under Grant No. 544937803 and was supported by several other grants in the past. The author further gratefully acknowledges funding from COST Action CA22133 PLANETS, supported by COST (European Cooperation in Science and Technology). The simulations were carried out on computing clusters hosted by the Max Planck Computing and Data Facility (MPCDF).

    During the preparation of this work, the author used Codex and Claude models as assistants for software development and manuscript preparation. Software-related assistance included code implementation and refactoring, debugging, and performance optimization. Manuscript-related assistance included literature searches, code--text cross-checking, and language revision. The author reviewed and validated all AI-assisted content and takes full responsibility for this publication.
\end{acknowledgements}

\bibliography{refs}
\bibliographystyle{styl/aa}

\begin{appendix} \nolinenumbers
\section{Analytical solutions to the Smoluchowski equation} \label{sec:smol}
\subsection{Constant kernel}

For the constant-kernel test, Eq.~(\ref{eq:kernel}) was replaced by
\begin{equation}
    K_{ij}\left(m_i, m_j\right)\Big|_{\rm const} = 1.
\end{equation}
All particles were initialized with $m_i = m_0$, such that the grain-number distribution satisfies
\begin{equation} \label{eq:mono_init}
    n\left(m, t = 0\right) = \frac{M_{\rm d}}{m_0^2}\delta_{\tilde{m}1},
\end{equation}
where $\delta_{\tilde{m}1}$ is the Kronecker delta and 
\begin{equation}
    \tilde{m} = m/m_0 = 1, 2, 3, \dots
\end{equation}
is the dimensionless mass, which remains a positive integer because of the discrete mass initialization in Eq.~(\ref{eq:mono_init}) and the coagulation-only prescription. The Smoluchowski equation then gives (\citealt{ohtsuki.K.1990.01})
\begin{equation} \label{eq:sol_const}
    n\left(m, t\right)\Big|_{\rm const}
        = \frac{M_{\rm d}}{m_0^2}\frac{
            \displaystyle\left(\tilde{t} / 2\right)^{\tilde{m} - 1}
        }{
            \displaystyle\left(1 + \tilde{t} / 2\right)^{\tilde{m} + 1}
        }.
\end{equation}
Notably, Eq.~(\ref{eq:sol_const}) offers
\begin{equation}
    m^2n\left(m, t\right)\Big|_{\rm const}
        \propto \tilde{m}^2\left(\frac{\tilde{t}}{2 + \tilde{t}}\right)^{\tilde{m} - 1},
\end{equation}
which has a maximum for $t > 0$ at
\begin{equation} \label{eq:m_peak}
    m_{\rm peak}(t) = -\frac{2m_0}{\ln\left(\displaystyle\frac{\tilde{t}}{2 + \tilde{t}}\right)}.
\end{equation}
This analytical form was used for comparison with the simulation results.

\subsection{Linear kernel}

For the linear-kernel test, Eq.~(\ref{eq:kernel}) was replaced by
\begin{equation}
    K_{ij}\left(m_i, m_j\right)\Big|_{\rm linear} = \frac{m_i + m_j}{m_0}.
\end{equation}
We sampled the grain masses of the particles from a gamma distribution with a shape parameter of 2 and a scale parameter of $m_0$, while retaining equal particle masses. This gives the initial grain-number distribution
\begin{equation}
    n\left(m, 0\right) = \frac{M_{\rm d}}{m_0^2}{\rm e}^{-\tilde{m}}.
\end{equation}
The Smoluchowski equation then gives (\citealt{ohtsuki.K.1990.01})
\begin{equation} \label{eq:sol_linear}
    \begin{aligned}
        n\left(m, t\right)\Big|_{\rm linear} & =
            \frac{1}{m}\frac{M_{\rm d}}{m_0}\frac{g}{\displaystyle\sqrt{1 - g}}
            \exp\left[-\tilde{m}\left(2 - g\right)\right] \\[4pt]
        & \times I_1\left(2\tilde{m}\sqrt{1 - g}\right),
    \end{aligned}
\end{equation}
where $I_1$ is the modified Bessel function of the first kind of order one and 
\begin{equation}
    g(t) = \exp\left(-\tilde{t}\right).
\end{equation}
To avoid overflow when evaluating the exponentially growing $I_1(x)$, we rewrite Eq.~(\ref{eq:sol_linear}) as
\begin{align}
    n\left(m, t\right)\Big|_{\rm linear} & =
        \frac{1}{m}\frac{M_{\rm d}}{m_0}\frac{g}{\sqrt{1 - g}}
        \exp\left[-\tilde{m}\left(\frac{g}{1 + \sqrt{1 - g}}\right)^2\right] \notag \\[2pt]
    & \times I_{1{\rm e}}\left(2\tilde{m}\sqrt{1 - g}\right),
\end{align}
where
\begin{equation}
    I_{1{\rm e}}(x) = {\rm e}^{-x}I_1(x)
\end{equation}
is the corresponding exponentially scaled Bessel function.

\subsection{Product kernel}

For the product-kernel test, Eq.~(\ref{eq:kernel}) was replaced by
\begin{equation}
    K_{ij}\left(m_i, m_j\right)\Big|_{\rm prod} = \frac{m_i \times m_j}{m_0^2}.
\end{equation}
With the initial distribution given by Eq.~(\ref{eq:mono_init}), the Smoluchowski equation gives (\citealt{trubnikov.BA.1971.08})
\begin{equation}
    n\left(m, t\right)\Big|_{\rm prod}
        = \frac{1}{m}\frac{M_{\rm d}}{m_0}
        \frac{\left(\tilde{m}\tilde{t}\right)^{\tilde{m} - 1}}{\tilde{m}!}
        \exp\left(-\tilde{m}\tilde{t}\right).
\end{equation}

\section{Parameter survey of the coagulation tests} \label{sec:survey}

\begin{figure*}[t]
\centering
\includegraphics[width = \textwidth]{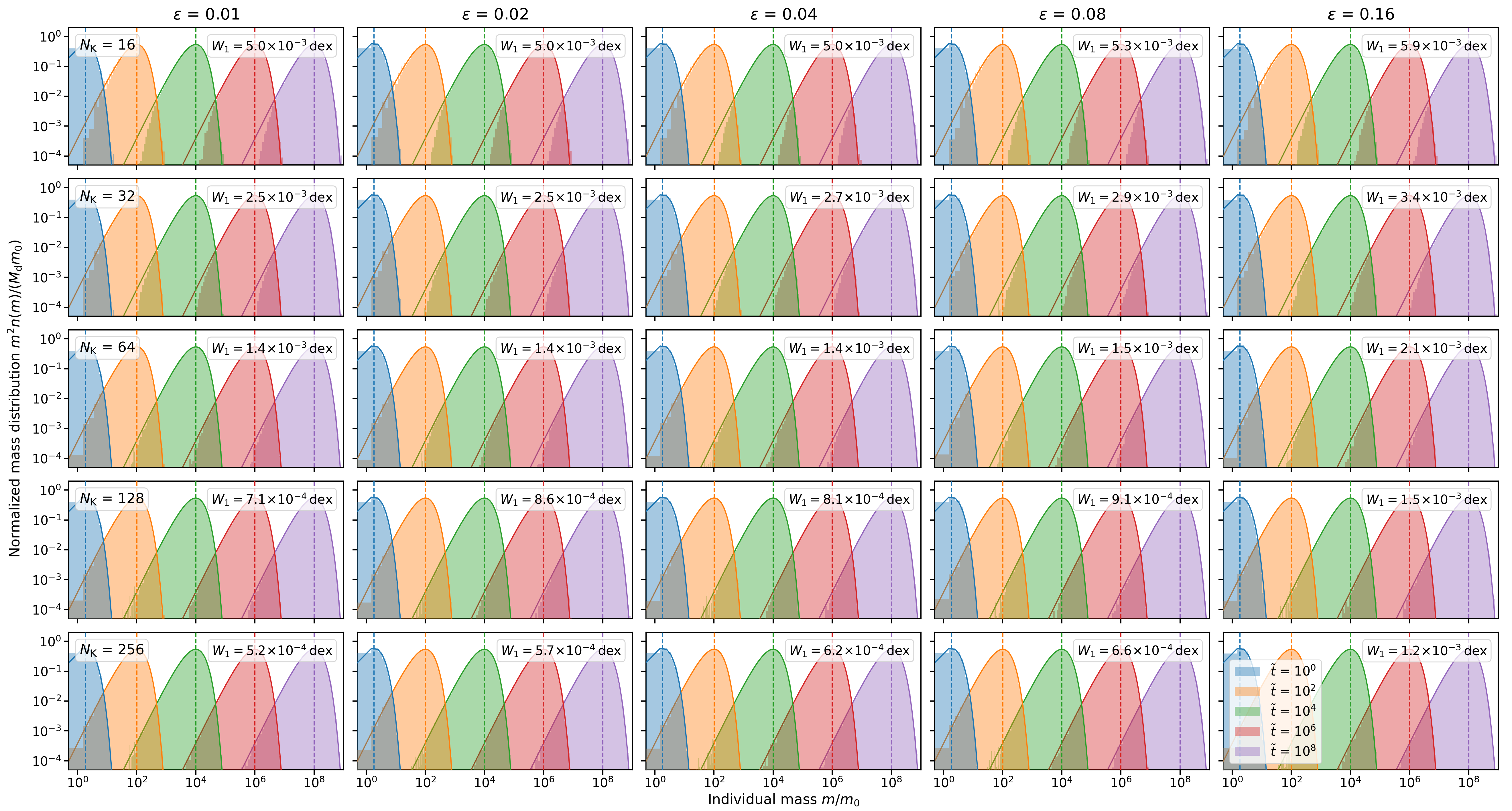}
\caption{
    Normalized grain-mass distributions in the constant-kernel test for all combinations of $N_{\rm K}$ (rows) and $\varepsilon$ (columns). Shaded histograms show one repetition and solid curves show the analytical solutions at $\tilde{t} = 10^0, 10^2, \dots, 10^8$. Vertical dashed lines mark the analytical peak masses $m_{\rm peak}$. Each panel lists $W_1$ at the final snapshot, averaged over ten repetitions.
\label{fig:coag_const}}
\end{figure*}

\begin{figure*}[t]
\centering
\includegraphics[width = \textwidth]{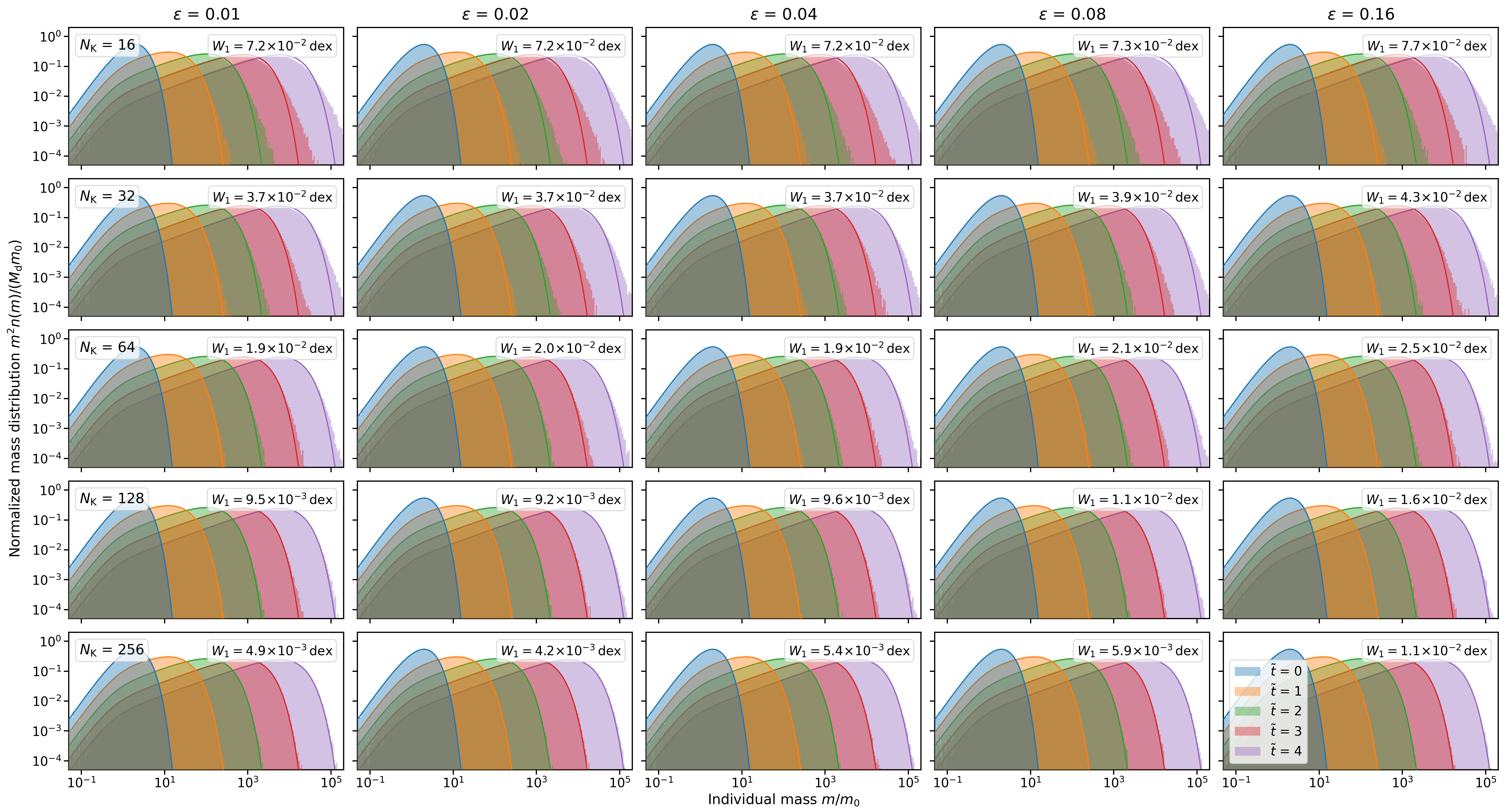}
\caption{
    Same as Fig.~\ref{fig:coag_const} but for the linear-kernel test at $\tilde{t} = 0, 1, \dots, 4$.
\label{fig:coag_linear}}
\end{figure*}

\begin{figure*}[t]
\centering
\includegraphics[width = \textwidth]{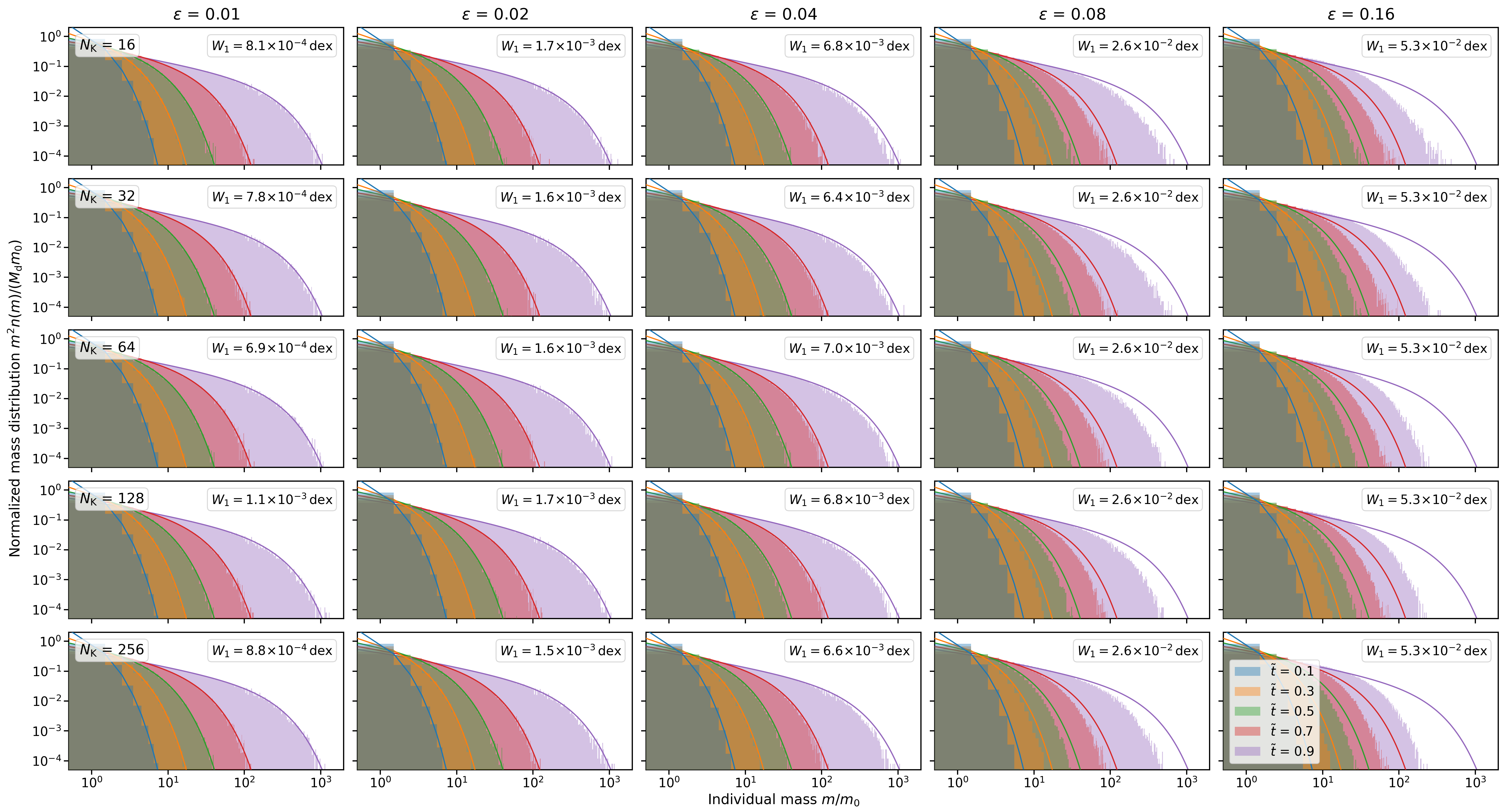}
\caption{
    Same as Fig.~\ref{fig:coag_const} but for the product-kernel test with neighbor reshuffling at $\tilde{t} = 0.1, 0.3, \dots, 0.9$.
\label{fig:coag_prod}}
\end{figure*}

\begin{figure}[t]
\centering
\includegraphics[width = \columnwidth]{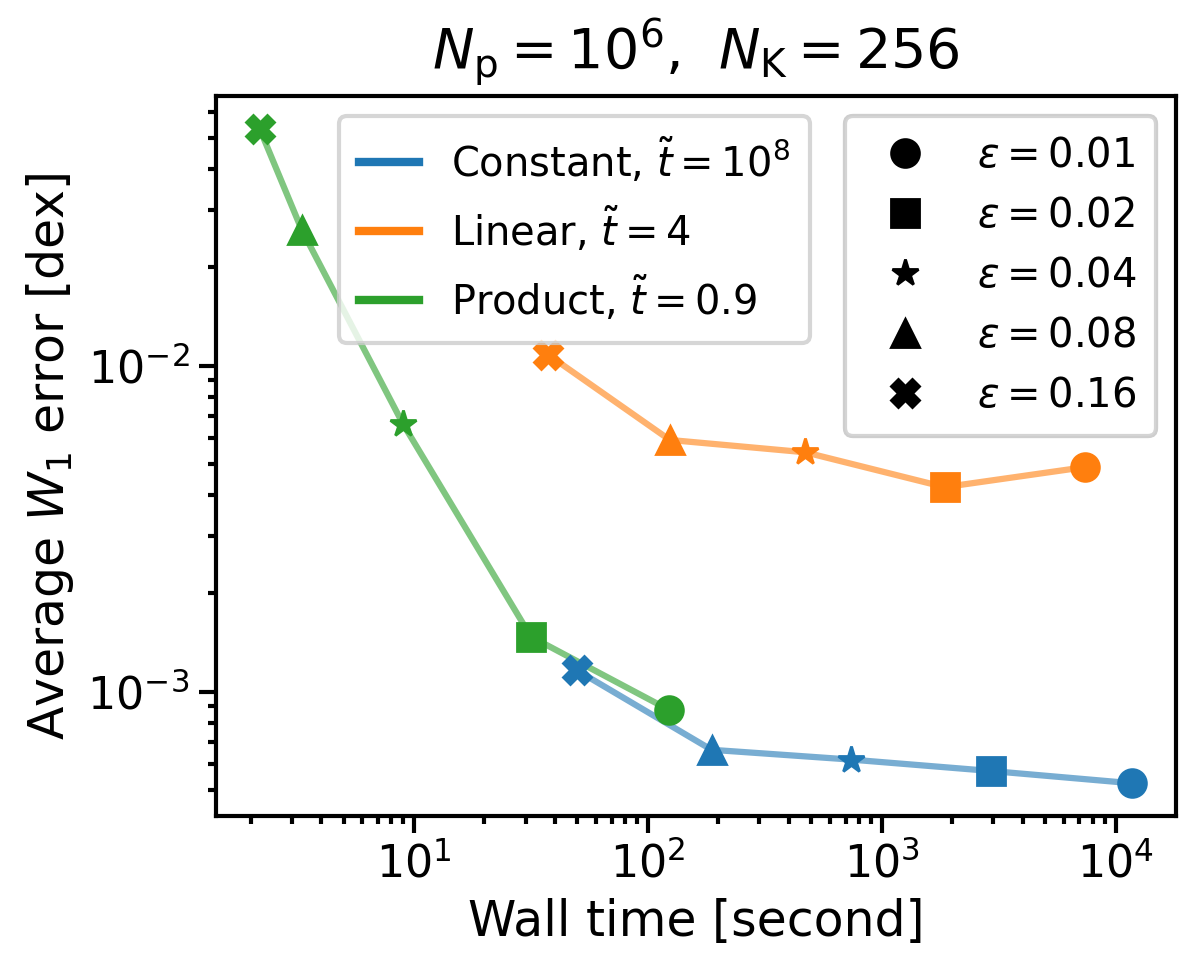}
\caption{
    Accuracy versus elapsed time of the coagulation tests with $N_{\rm K} = 256$. Colors denote the collision kernels and markers denote $\varepsilon$. Both $W_1$ at the final snapshot and the elapsed time are averaged over ten repetitions.
\label{fig:coag_cost}}
\end{figure}

For each collision kernel, we ran all 25 combinations of $N_{\rm K}$ and $\varepsilon$ listed in Sect.~\ref{sec:test_coag}, each with ten independent repetitions that differed in the random numbers used for particle positions and collision sampling, as well as the initial grain masses for the linear-kernel tests. The constant-kernel models were evolved to $\tilde{t} = 10^8$, the linear-kernel models to $\tilde{t} = 4$, and the product-kernel models to $\tilde{t} = 0.9$, before gelation at $\tilde{t} = 1$. Figures~\ref{fig:coag_const}, \ref{fig:coag_linear}, and \ref{fig:coag_prod} compare the grain-mass distributions of the repetition with seed index 0 with the analytical solutions in Appendix~\ref{sec:smol}.

We quantified the discrepancy at the final snapshot using the first Wasserstein distance $W_1$ between the simulated and analytical mass-weighted distributions in logarithmic grain mass \citep{vallender.SS.1974.09}
\begin{equation}
    W_1 = \int\left|
        F_{\rm sim}\left(\tilde{m}\right) - F_{\rm exact}\left(\tilde{m}\right)
    \right|d\log_{10}\tilde{m},
\end{equation}
where $F_{\rm sim}\left(\tilde{m}\right)$ and $F_{\rm exact}\left(\tilde{m}\right)$ are the simulated and analytical fractions of the total dust mass in grains with dimensionless mass below $\tilde{m}$, and the integral covers the range occupied by both distributions. The distance $W_1$ is the mass-weighted mean shift in $\log_{10}m$ required to transform one distribution into the other, such that $W_1 = 0.01$~dex corresponds to a typical mass offset of about 2\%. The reported values of $W_1$ are averages over the ten repetitions, whose standard deviation is typically $\lesssim 25\%$ of the mean for the constant and linear kernels and up to $\sim 60\%$ for the product kernel at $\varepsilon \leq 0.02$, where $W_1$ is smallest.

For the product kernel, the collision rates are dominated by the few most massive grains. With fixed neighbor lists, a growing particle repeatedly samples the same partners, which correlates the growth histories of nearby particles and can artificially amplify runaway growth. Because the analytical solution assumes a well-mixed population, we randomly permuted the cached neighbor indices of all particles before every neighborhood refresh in the product-kernel models, and treated the domain as a single spatial group so that all particles refreshed simultaneously. This reshuffling was applied only in the product-kernel test. It removes the dependence of the results on $N_{\rm K}$ (Fig.~\ref{fig:coag_prod}), so that this test primarily probes the frozen-neighbor intervals controlled by $\varepsilon$.

The wall time was measured as the elapsed time of each entire simulation, including initialization and output. Because the particles did not move, each collision timestep spanned an output interval, and the KD-tree and the neighbor search were constructed only once. The wall time is therefore dominated by the collisional evolution and underestimates the cost of models with transport, in which the neighbor search is repeated every timestep. For the product kernel, the wall time also includes the reshuffling overhead.

Figure~\ref{fig:coag_cost} summarizes the accuracy and wall time of the models with $N_{\rm K} = 256$. For the constant and linear kernels, $W_1$ depends only weakly on $\varepsilon$ up to $\varepsilon = 0.08$ and decreases with increasing $N_{\rm K}$, approximately as $N_{\rm K}^{-1}$ for the linear kernel and as $N_{\rm K}^{-0.8}$ for the constant kernel (Figs.~\ref{fig:coag_const} and \ref{fig:coag_linear}), indicating that the error is dominated by the finite neighborhood sampling rather than by the frozen-neighbor intervals. In contrast, the product-kernel error increases rapidly with $\varepsilon$: at large $\varepsilon$, the simulated distributions fall below the analytical solutions at the high-mass end. This is consistent with partner masses cached at the beginning of each interval lagging behind their actual growth, to which the product kernel is most sensitive. For the constant and linear kernels, the wall time scales approximately as $\varepsilon^{-2}$, consistent with the second-moment constraint in Sect.~\ref{sec:mc} limiting the frozen-neighbor intervals.

\section{Elapsed time in the benchmark comparison}

\begin{figure*}[t]
\centering
\includegraphics[width = \textwidth]{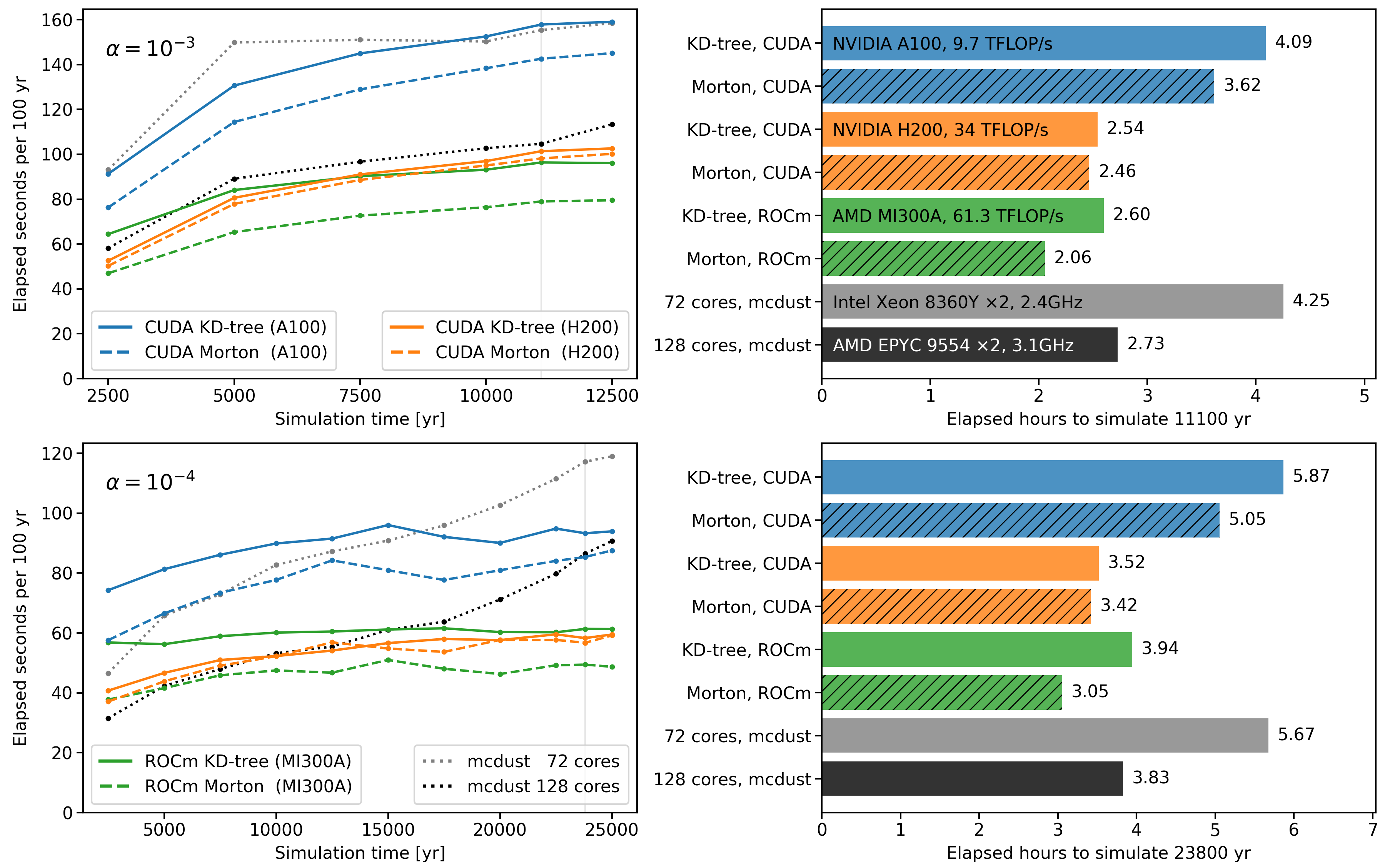}
\caption{
    Elapsed computational time of \texttt{GameDev} and \texttt{mcdust} for the \citetalias{eriksson.LEJ.2026.07} benchmarks with $\alpha = 10^{-3}$ (top) and $\alpha = 10^{-4}$ (bottom). Left: elapsed time per 100~yr of simulation time. Right: total elapsed time to reach the snapshots shown in Figs.~\ref{fig:e26_1} and \ref{fig:e26_2}. Labels give the hardware used by each model, with the peak double-precision vector throughput for the GPUs and the base clock frequency for the CPUs.
\label{fig:perf}}
\end{figure*}

Figure~\ref{fig:perf} shows the elapsed computational time of the \texttt{GameDev} and \texttt{mcdust} models discussed in Sect.~\ref{sec:perf}.

\section{The dust fluid model} \label{sec:fluid}

In addition to the particle model, \texttt{GameDev} provides a standalone Eulerian model that treats dust as a pressureless fluid. The fluid model shares the prescriptions of aerodynamic drag, radiation pressure, diffusion, and the analytical gas background with the particle model, and we describe only the differences here.

\subsection{Governing equations} \label{sec:fluid_eq}

The fluid model represents a single dust species with a fixed grain size, so that the Stokes number depends only on position. It supports the 3D and the 2D radial--azimuthal configurations, both of which require a resolved azimuthal dimension. Collisions and Poynting--Robertson drag are not included. With the dust density $\varrho_{\rm d}$ defined in Sect.~\ref{sec:geom}, the dust evolves according to
\begin{equation}
    \frac{\partial\varrho_{\rm d}}{\partial t} + \nabla\cdot\left(\varrho_{\rm d}\boldsymbol{v}\right)
        = \left.\frac{\partial\varrho_{\rm d}}{\partial t}\right|_{\rm diffusion}
\end{equation}
and
\begin{equation} \label{eq:fluid_mom}
    \begin{aligned}
        \frac{\partial\left(\varrho_{\rm d}\boldsymbol{v}\right)}{\partial t}
        & + \nabla\cdot\left(\varrho_{\rm d}\boldsymbol{v}\boldsymbol{v}\right) = \\[4pt]
        & \varrho_{\rm d}\left(
            -\frac{GM_\star}{r^2}\hat{\boldsymbol{r}} + \boldsymbol{f}_{\rm RP}
            + \frac{\boldsymbol{v}_{\rm g} - \boldsymbol{v}}{t_{\rm s}}
        \right) + \boldsymbol{S}_{\rm diff},
    \end{aligned}
\end{equation}
where $\boldsymbol{v}$ is the dust velocity, the diffusion term is given by Eq.~(\ref{eq:diff_conc}) or (\ref{eq:diff_dens}), and $\boldsymbol{S}_{\rm diff}$ is the momentum flux carried by the diffusive mass flux (Appendix~\ref{sec:fluid_num}). Given that dust is pressureless, the momentum flux $\varrho_{\rm d}\boldsymbol{v}\boldsymbol{v}$ does not include the velocity-dispersion term. This closure is appropriate as long as the dust velocity is locally single-valued. Unlike the particle model, however, the fluid model cannot represent multiple velocity streams at the same location after trajectories cross, as can occur for weakly coupled grains.

Same as the particle model, the momentum is evolved in terms of $\boldsymbol{y} = \left(v_r, \ell_\theta, \ell_\phi\right)$, and the conserved variables are $\varrho_{\rm d}$ and $\varrho_{\rm d}\boldsymbol{y}$, with $\ell_\theta$ omitted in 2D. The dust diffusion coefficients keep the form given in Sect.~\ref{sec:diff}, with a Schmidt number in each direction, but the diffusion tensor is diagonal in the local spherical basis instead of the cylindrical basis. The two bases coincide in vertically integrated models, whereas in 3D the fluid model diffuses along $r$ rather than $R$.

\subsection{Numerical methods} \label{sec:fluid_num}

Density and momenta are transported conservatively by directionally split finite-volume sweeps. Interface states are reconstructed with the piecewise parabolic method (PPM; \citealt{colella.P.1984.09}). Fluxes are calculated with the Harten--Lax--van--Leer method (HLL; \citealt{harten.A.1983.01}). Given that the pressureless equations are only weakly hyperbolic and can develop vacuum regions and singular concentrations, the high-order flux at each face is blended with a first-order HLL flux such that the density remains nonnegative and the transported specific momenta remain within local bounds (\citealt{guermond.JL.2019.04}). Azimuthal transport uses the fast advection in rotating gaseous objects method (FARGO; \citealt{masset.FS.2000.01}). Radial and polar transport are integrated with the three-stage third-order strong stability preserving Runge--Kutta method (SSPRK(3,3); \citealt{shu.CW.1988.08}). The radial and polar boundaries allow only outflow, and the azimuthal boundaries are periodic.

Stellar gravity, aerodynamic drag, and radiation pressure acting on $\boldsymbol{y}$ are integrated locally in each cell. Same as the particle model, drag relaxation is integrated analytically for a fixed stopping time, while the remaining forces are interpolated linearly in time between their old and updated values with drag-dependent weights, which recover the trapezoidal rule for $\Delta t \ll t_{\rm s}$. The optical depth is calculated from the cell densities following Eq.~(\ref{eq:tau_3d}) or (\ref{eq:tau_2d}).

Dust diffusion is solved in finite-volume form with the implicit Crank--Nicolson method (\citealt{crank.J.1947.01}) along each direction, using subcycling and donor-limited fluxes where necessary to preserve positivity. To keep the momentum consistent with the redistributed mass, the diffusive mass flux across each face carries the specific momentum $\boldsymbol{y}$ of its donor cell, which defines $\boldsymbol{S}_{\rm diff}$ in Eq.~(\ref{eq:fluid_mom}) and conserves momentum across internal faces.

Similar to the particle model, each timestep follows a symmetric operator splitting,
\begin{equation*}
    \mathcal{D}\left(\frac{\Delta t}{2}\right)
    \mathcal{T}\left(\frac{\Delta t}{2}\right)
    \mathcal{S}\left(\vphantom{\frac{\Delta t}{2}}\Delta t\right)
    \mathcal{T}\left(\frac{\Delta t}{2}\right)
    \mathcal{D}\left(\frac{\Delta t}{2}\right),
\end{equation*}
where $\mathcal{D}$, $\mathcal{T}$, and $\mathcal{S}$ are diffusion, transport, and local source terms, respectively. The directional sweeps within $\mathcal{D}$ and $\mathcal{T}$ are applied in reverse order in the second half of the step. The optical depth is evaluated at the intermediate state before $\mathcal{S}$. The timestep is limited by the azimuthal velocity relative to the ring-averaged orbital motion removed by FARGO and by the radial and polar velocities relative to the local cell sizes. The transport half-steps are subcycled whenever this limit becomes more restrictive during the step. The implicit diffusion and the analytical drag integration impose no additional timestep constraints.

\subsection{Application to the photospheric instability} \label{sec:fluid_psi}

\begin{figure}[t]
\centering
\includegraphics[width = \columnwidth]{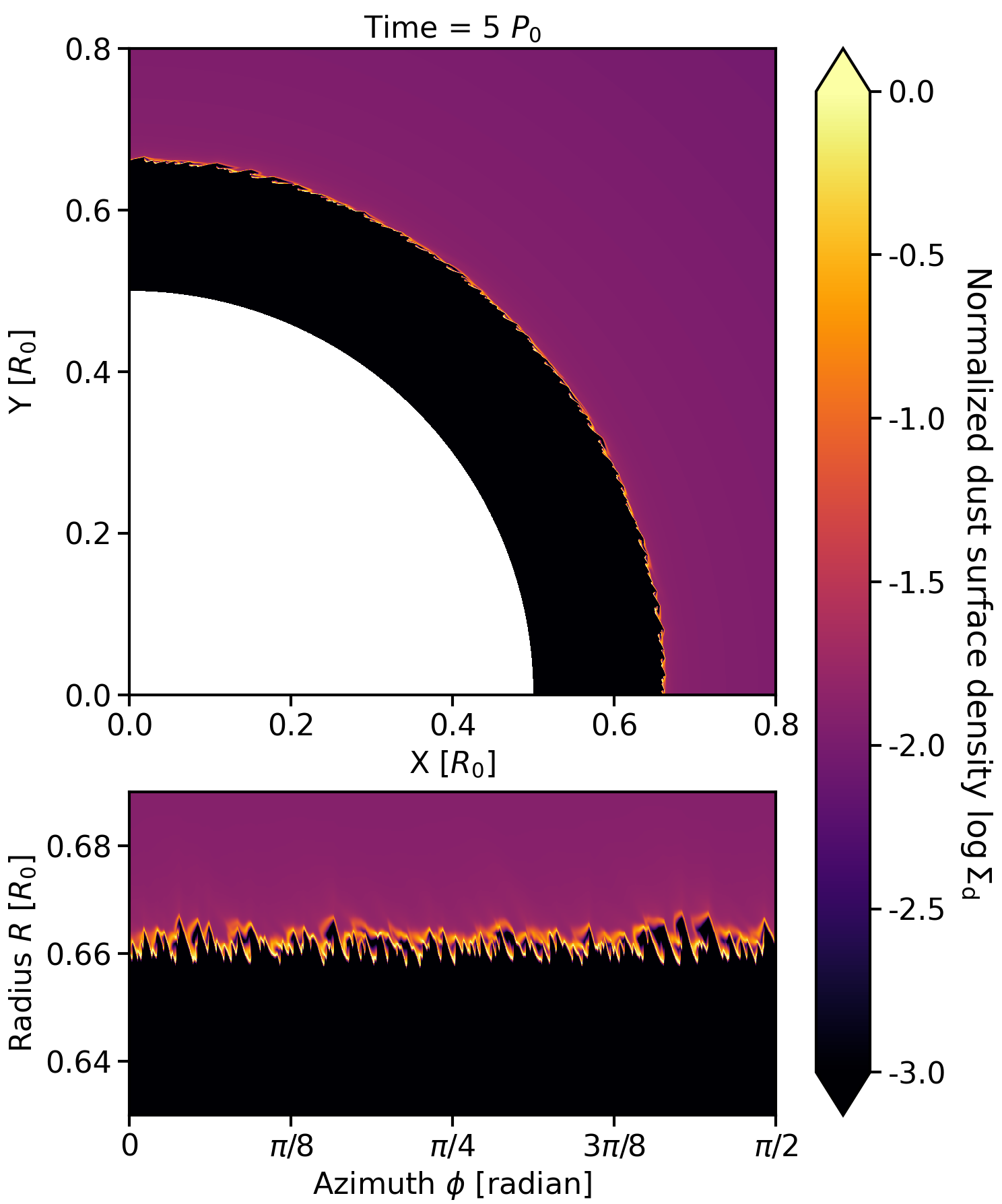}
\caption{
    Same as Fig.~\ref{fig:psi} but for the fluid model at $t = 5P_0$.
\label{fig:psi_fluid}}
\end{figure}

The fluid model was primarily motivated by studies of the linear-growth phase of the PSI that require extremely weak initial perturbations. In the particle model, the initial perturbation is set by the particle-sampling Poisson noise, which decreases only as $N_{\rm p}^{-1/2}$. Reducing it from $\sim 3\%$ to $10^{-10}$ on the same grid would require $\sim 10^{26}$ particles, far beyond what is computationally feasible.

We repeated the model of Sect.~\ref{sec:psi} with the fluid model, using the same physical parameters and initial surface density on a $3072\times4096$ grid in the radial and azimuthal directions. Instead of sampling noise, the initial surface density was perturbed in azimuth with a relative amplitude of $10^{-10}$, drawn independently from a standard normal distribution for each azimuthal cell and applied identically at all radii. Because the instability likely requires longer to grow from this weaker seed, Fig.~\ref{fig:psi_fluid} shows a later snapshot at $t = 5P_0$, compared with $t = 3P_0$ for the particle model in Fig.~\ref{fig:psi}. The fluid model develops a clumpy disk edge similar to that of the particle model.

\clearpage
\onecolumn
\section{List of symbols}

Table~\ref{tab:symbols} lists the symbols used in this paper. Bold italic symbols denote vectors, bold upright symbols denote matrices and tensors, and calligraphic symbols denote sets, operators, and probability distributions. The subscript 0 denotes reference values, the subscript ``exact'' denotes analytical solutions, and a tilde denotes a quantity normalized by its reference value.

\normalsize
\renewcommand{\arraystretch}{1.15}
\begin{longtable}{l p{11.5cm} l}
\caption{Symbols used in this paper.}
\label{tab:symbols} \\
\hline\hline
\rule[-0.4125\baselineskip]{0pt}{1.375\baselineskip}Symbol & Description & Section \\
\hline
\endfirsthead
\caption{continued.} \\
\hline\hline
\rule[-0.4125\baselineskip]{0pt}{1.375\baselineskip}Symbol & Description & Section \\
\hline
\endhead
\hline
\endfoot
\multicolumn{3}{l}{\rule[-0.4125\baselineskip]{0pt}{1.375\baselineskip}\textit{Latin letters}} \\
\hline
$a$ & Grain radius, $s/2$ & \ref{sec:test_e26} \\
$A_b$ & First moment of the grain-size change rate in controller bin $b$ & \ref{sec:mc} \\
$b$ & Controller bin index & \ref{sec:mc} \\
$B_b$ & Second moment of the grain-size change rate in controller bin $b$ & \ref{sec:mc} \\
$c$ & Speed of light & \ref{sec:rad} \\
$c_{\rm s}$, $c_{\rm s0}$ & Sound speed, and its value at $R_0$ & \ref{sec:gas}, \ref{sec:drag} \\
$C$ & Adaptive safety factor of the frozen-neighbor intervals & \ref{sec:mc} \\
$d_i$ & Distance to the farthest retained neighbor of particle $i$ & \ref{sec:rate} \\
$D$, $D_0$ & Dust diffusion coefficient, and its value at $R_0$ & \ref{sec:diff}, \ref{sec:test_diff} \\
$\mathbf{D}$ & Dust diffusion tensor & \ref{sec:diff} \\
$D_R$, $D_Z$, $D_\phi$ & Directional dust diffusion coefficients & \ref{sec:diff} \\
$e$ & Orbital eccentricity & \ref{sec:test_orbit} \\
$E$, $E_{\rm exact}$ & Specific orbital energy, and its exact value & \ref{sec:test_orbit} \\
$\boldsymbol{f}$, $f_r$, $f_\theta$, $f_\phi$ & Remaining specific force terms in the equation of $\boldsymbol{y}$, and their components & \ref{sec:ssa} \\
$\boldsymbol{f}_{\rm aero}$ & Specific aerodynamic force & \ref{sec:dyn} \\
$\boldsymbol{f}_{\rm PR}$ & Specific Poynting--Robertson drag & \ref{sec:rad} \\
$\boldsymbol{f}_{\rm rad}$ & Specific radiation force & \ref{sec:dyn} \\
$\boldsymbol{f}_{\rm RP}$ & Specific radiation pressure & \ref{sec:rad} \\
$F_{\rm exact}$, $F_{\rm sim}$ & Analytical and simulated cumulative mass fractions & \ref{sec:survey} \\
$g$ & $\exp\left(-\tilde{t}\right)$ in the linear-kernel solution & \ref{sec:smol} \\
$G$ & Gravitational constant & \ref{sec:gas} \\
$h_{\rm g}$, $h_{\rm g0}$ & Gas aspect ratio, and its value at $R_0$ & \ref{sec:gas} \\
$H_{\rm g}$ & Gas scale height & \ref{sec:gas} \\
$H_i$, $H_j$ & Gas scale heights at particles $i$ and $j$, $H_{\rm g}\left(R_i\right)$ and $H_{\rm g}\left(R_j\right)$ & \ref{sec:rate} \\
$i$ & Query particle index & \ref{sec:particle} \\
$I_1$, $I_{1{\rm e}}$ & Modified Bessel function of the first kind of order one, and its exponentially scaled form & \ref{sec:smol} \\
$j$ & Neighbor (partner population) index & \ref{sec:rate} \\
$J_{io}$, $\left\langle J_{io}\right\rangle$ & Logarithmic grain-size change for outcome $o$, and its expectation & \ref{sec:mc} \\
$k_{\rm B}$ & Boltzmann constant & \ref{sec:outcome} \\
$K_{ij}$ & Collision kernel & \ref{sec:rate} \\
$\ell_\theta$, $\ell_\phi$ & Specific polar and azimuthal angular momenta & \ref{sec:ssa} \\
$L_\star$ & Stellar luminosity & \ref{sec:rad} \\
$m$, $m_i$, $m_j$, $m_0$ & Grain mass, and the unit (monomer) mass & \ref{sec:particle}, \ref{sec:test_coag} \\
$m_{\rm peak}$ & Grain mass at the peak of $m^2n$ & \ref{sec:smol} \\
$m_i^\prime$ & Post-collision grain mass & \ref{sec:outcome} \\
$\tilde{m}$ & Dimensionless grain mass, $m/m_0$ & \ref{sec:smol} \\
$M_{\rm d}$ & Total dust mass & \ref{sec:test_diff}, \ref{sec:test_coag} \\
$M_{\rm frag}$ & Mass-weighted fragment distribution & \ref{sec:outcome} \\
$M_i$ & Particle mass & \ref{sec:particle} \\
$M_\star$ & Stellar mass & \ref{sec:gas} \\
$n$ & Grain-number distribution per unit mass & \ref{sec:smol} \\
$n_{ij}$ & Number density of grains represented by neighbor $j$ & \ref{sec:rate} \\
$N_{\rm coll}$ & Grouping factor (number of grouped collisions) & \ref{sec:outcome} \\
$N_{\rm frag}$ & Number-weighted fragment distribution & \ref{sec:outcome} \\
$N_i$ & Number of physical dust grains represented by particle $i$ & \ref{sec:particle} \\
$N_{\rm K}$ & Number of nearest neighbors & \ref{sec:rate} \\
$N_{\rm p}$ & Number of representative particles & \ref{sec:particle} \\
$N_i^\prime$ & Post-collision number of represented grains & \ref{sec:outcome} \\
$o$ & Collision outcome index & \ref{sec:mc} \\
$p$ & Power-law index of the gas surface density & \ref{sec:gas} \\
$P_0$ & Orbital period at $R_0$, $2\pi/\Omega_{\rm K0}$ & \ref{sec:test_orbit} \\
$q$ & Power-law index of the gas temperature & \ref{sec:gas} \\
$r$ & Spherical radius & \ref{sec:geom} \\
$r_{\rm min}$ & Inner boundary in spherical radius & \ref{sec:rad} \\
$\hat{\boldsymbol{r}}$ & Radial unit vector & \ref{sec:dyn} \\
$R$, $R_0$ & Cylindrical radius, and the reference radius & \ref{sec:geom}, \ref{sec:gas} \\
$R_{\rm min}$ & Inner boundary in cylindrical radius & \ref{sec:rad} \\
$\Delta R$ & Diffusive radial displacement & \ref{sec:diff} \\
$s$, $s_i$, $s_j$, $s_0$ & Grain diameter, and the reference grain diameter & \ref{sec:particle}, \ref{sec:drag} \\
$s_{\rm min}$ & Minimum grain diameter of fragments & \ref{sec:outcome} \\
$s_i^\prime$ & Post-collision grain diameter & \ref{sec:outcome} \\
$\boldsymbol{S}_{\rm diff}$ & Momentum flux carried by the diffusive mass flux & \ref{sec:fluid_eq} \\
${\rm Sc}$ & Schmidt number of the gas turbulence, relating $\nu$ to $D$ & \ref{sec:diff} \\
${\rm St}$, ${\rm St}_0$ & Stokes number, and its midplane value for $s_0$ at $R_0$ & \ref{sec:drag} \\
$t$, $t_0$ & Time, and the reference time & \ref{sec:dyn}, \ref{sec:test_diff}, \ref{sec:test_coag} \\
$t_{\rm s}$ & Stopping time & \ref{sec:drag} \\
$t_\beta$ & Timescale of the radiation taper & \ref{sec:rad} \\
$\tilde{t}$ & Dimensionless time, $t/t_0$ & \ref{sec:test_diff}, \ref{sec:test_coag} \\
$\Delta t$ & Timestep, over which the transport operator advances & \ref{sec:ssa}, \ref{sec:step} \\
$\Delta t_{\rm coll}$ & Collision timestep, $\Delta t/2$ in models with transport & \ref{sec:mc}, \ref{sec:step} \\
$\Delta t_{\rm diff}$ & Diffusion timestep, $\Delta t/2$ & \ref{sec:diff}, \ref{sec:step} \\
$\Delta t_{\rm fn}$ & Frozen-neighbor interval & \ref{sec:mc} \\
$\Delta t_{\rm fn,exe}$ & Executed frozen-neighbor interval & \ref{sec:mc} \\
$\Delta t_{\rm fn,max}$ & Maximum frozen-neighbor interval & \ref{sec:mc} \\
$\Delta t_{\rm fn,req}$ & Requested frozen-neighbor interval & \ref{sec:mc} \\
$\Delta t_i$ & Most restrictive local timestep of particle $i$ & \ref{sec:step} \\
$\Delta t_{\rm max}$ & Maximum transport timestep & \ref{sec:step} \\
$\Delta t_{\rm out}$ & Remaining time before the next output & \ref{sec:step} \\
$\Delta t_{\rm remain}$ & Remaining time in the current collision timestep $\Delta t_{\rm coll}$ & \ref{sec:mc} \\
$\Delta t_{\rm w}$ & Waiting time until the next collision event & \ref{sec:mc} \\
$T$ & Gas temperature & \ref{sec:outcome} \\
$u$ & Eccentric anomaly & \ref{sec:test_orbit} \\
$U$, $U_i$ & Uniform random number & \ref{sec:outcome}, \ref{sec:mc} \\
$\boldsymbol{v}$, $v_r$, $v_\theta$, $v_\phi$ & Dust velocity and its spherical components & \ref{sec:particle}, \ref{sec:fluid_eq} \\
$v_{\rm frag}$ & Fragmentation speed & \ref{sec:outcome} \\
$\boldsymbol{v}_{\rm g}$ & Gas velocity & \ref{sec:drag} \\
$v_{{\rm g},\phi}$ & Gas azimuthal velocity & \ref{sec:gas} \\
$v_{\rm visc}$ & Viscous radial gas velocity & \ref{sec:gas} \\
$\Delta v_{ij}$ & Collision speed & \ref{sec:outcome} \\
$\Delta v_{\rm B}$ & Brownian contribution to the collision speed & \ref{sec:outcome} \\
$\Delta v_R$, $\Delta v_Z$, $\Delta v_\phi$ & Relative radial drift, vertical settling, and azimuthal speeds & \ref{sec:outcome} \\
$\Delta v_{\rm t}$ & Turbulent contribution to the collision speed & \ref{sec:outcome} \\
$V_i$ & Sampling measure of particle $i$ (volume, area, or annulus area) & \ref{sec:rate} \\
$w$ & Width of the Gaussian smoothing kernel & \ref{sec:psi} \\
$W_1$ & Wasserstein distance & \ref{sec:survey} \\
$\boldsymbol{x}$, $\boldsymbol{x}_i$ & Position vector & \ref{sec:particle} \\
$\boldsymbol{y}$ & Integration variables $\left(v_r, \ell_\theta, \ell_\phi\right)$ & \ref{sec:ssa} \\
$\boldsymbol{y}_{\rm g}$ & Gas velocity in the integration variables & \ref{sec:ssa} \\
$Z$ & Cylindrical height & \ref{sec:geom} \\
$\Delta Z$ & Diffusive vertical displacement & \ref{sec:diff} \\
\hline
\multicolumn{3}{l}{\rule[-0.4125\baselineskip]{0pt}{1.375\baselineskip}\textit{Greek letters}} \\
\hline
$\alpha$ & Viscosity parameter & \ref{sec:gas} \\
$\beta$, $\beta_0$ & Ratio of radiation pressure to gravity, and its value for $s_0$ & \ref{sec:rad} \\
$\gamma_{\rm PR}$ & Poynting--Robertson damping rate & \ref{sec:rad} \\
$\gamma_r$, $\gamma_\theta$, $\gamma_\phi$ & Damping rates of the integration variables & \ref{sec:ssa} \\
$\mathbf{\Gamma}$ & Diagonal damping-rate matrix & \ref{sec:ssa} \\
$\delta_{\tilde{m}1}$ & Kronecker delta & \ref{sec:smol} \\
$\varepsilon$ & Tolerance of the frozen-neighbor intervals & \ref{sec:mc} \\
$\zeta_R$, $\zeta_Z$, $\zeta_\phi$ & Independent standard normal random numbers & \ref{sec:diff} \\
$\eta$ & Radial pressure-gradient parameter & \ref{sec:gas} \\
$\theta$ & Polar angle & \ref{sec:geom} \\
$\hat{\boldsymbol{\theta}}$ & Polar unit vector & \ref{sec:rad} \\
$\kappa$, $\kappa_0$ & Dust opacity, and its value for $s_0$ & \ref{sec:rad} \\
$\lambda_0$ & Collision-rate normalization & \ref{sec:test_coag} \\
$\lambda_i$, $\lambda_{ij}$ & Total and pair collision rates & \ref{sec:rate} \\
$\lambda^*_i$, $\lambda^*_{ij}$ & Effective total and pair collision rates & \ref{sec:outcome}, \ref{sec:mc} \\
$\lambda^*_{ij,{\rm deb}}$, $\lambda^*_{ij,{\rm rem}}$ & Effective rates of the erosion debris and remnant branches & \ref{sec:outcome} \\
$\mu$ & Mean of $\ln\left(R_i/R_0\right)$ & \ref{sec:test_diff} \\
$\nu$ & Gas kinematic viscosity & \ref{sec:gas} \\
$\xi$ & Argument of the radiation taper & \ref{sec:rad} \\
$\rho_{\rm d}$ & Dust volume density & \ref{sec:geom} \\
$\rho_{\rm g}$, $\rho_{\rm g0}$ & Gas volume density, and its midplane value at $R_0$ & \ref{sec:gas}, \ref{sec:drag} \\
$\rho_\circ$ & Internal density of grains & \ref{sec:particle} \\
$\varrho_{\rm d}$ & Dust density, $\Sigma_{\rm d}$ in vertically integrated models and $\rho_{\rm d}$ otherwise & \ref{sec:geom} \\
$\varrho_{\rm g}$ & Gas density, $\Sigma_{\rm g}$ in vertically integrated models and $\rho_{\rm g}$ otherwise & \ref{sec:geom} \\
$\sigma$ & Standard deviation of $\ln\left(R_i/R_0\right)$ & \ref{sec:test_diff} \\
$\Sigma_{\rm d}$, $\Sigma_{\rm d0}$ & Dust surface density, and the normalization of its initial profile & \ref{sec:geom}, \ref{sec:psi} \\
$\Sigma_{\rm g}$, $\Sigma_{\rm g0}$ & Gas surface density, and its value at $R_0$ & \ref{sec:gas} \\
$\tau$ & Radial optical depth & \ref{sec:rad} \\
$\phi$ & Azimuthal angle & \ref{sec:geom} \\
$\phi_{\rm exact}$ & Exact true anomaly & \ref{sec:test_orbit} \\
$\hat{\boldsymbol{\phi}}$ & Azimuthal unit vector & \ref{sec:rad} \\
$\Delta\phi$ & Diffusive azimuthal displacement & \ref{sec:diff} \\
$\chi$ & Switch between concentration ($\chi = 1$) and density ($\chi = 0$) diffusion & \ref{sec:diff} \\
$\psi$ & Grain-mass ratio, $m_j/m_i$ & \ref{sec:outcome} \\
$\Omega_{\rm K}$, $\Omega_{\rm K0}$ & Keplerian angular velocity, and its value at $R_0$ & \ref{sec:gas}, \ref{sec:drag} \\
\hline
\multicolumn{3}{l}{\rule[-0.4125\baselineskip]{0pt}{1.375\baselineskip}\textit{Calligraphic symbols}} \\
\hline
$\mathcal{B}$ & Set of controller bins in a spatial group & \ref{sec:mc} \\
$\mathcal{C}$ & Collision operator & \ref{sec:step} \\
$\mathcal{D}$ & Diffusion operator & \ref{sec:step}, \ref{sec:fluid_num} \\
$\mathcal{J}_i$ & Set of retained neighbors of particle $i$ & \ref{sec:rate} \\
$\mathcal{N}$ & Normal distribution & \ref{sec:diff}, \ref{sec:rate} \\
$\mathcal{S}$ & Local source operator of the fluid model & \ref{sec:fluid_num} \\
$\mathcal{T}$ & Deterministic transport operator & \ref{sec:step}, \ref{sec:fluid_num} \\
$\mathcal{U}$ & Uniform distribution & \ref{sec:outcome} \\
\end{longtable}
\renewcommand{\arraystretch}{1}
\normalsize

\end{appendix}

\end{document}